\documentclass{emulateapj}

\begin{document}

\title{Calibrating Ultracool Dwarfs: Optical Template Spectra, Bolometric Corrections, and $\chi$ Values $^{\dagger}$}

\author{Sarah J. Schmidt\altaffilmark{*,1,2}, Andrew A. West\altaffilmark{3}, John J. Bochanski\altaffilmark{4},
Suzanne L. Hawley\altaffilmark{2} Collin Kielty\altaffilmark{2},}

\altaffiltext{*} {schmidt@astronomy.ohio-state.edu}
\altaffiltext{1} {Department of Astronomy, Ohio State University, 140 West 18th Avenue, Columbus, OH 43210}
\altaffiltext{2} {Department of Astronomy, University of Washington, Box 351580, Seattle, WA 98195}
\altaffiltext{3} {Department of Astronomy, Boston University, CAS 422A, 725 Commonwealth Ave, Boston, MA 02215}
\altaffiltext{4} {Physics and Astronomy Department, Haverford College, 370 Lancaster Avenue, Haverford, PA 19041}
\footnotetext[$\dagger$]{This publication is partially based on observations obtained with the Apache Point Observatory 3.5-meter telescope, which is owned and operated by the Astrophysical Research Consortium.}

\begin{abstract}
We present optical template spectra, bolometric corrections, and $\chi$ values for ultracool dwarfs. The templates are based on spectra from the Sloan Digital Sky Survey (SDSS) and the Astrophysical Research Consortium 3.5-m telescope. The spectral features and overall shape of the L dwarf templates are consistent with previous spectroscopic standards and the templates have a radial velocity precision of $\sim$10--20 km s$^{-1}$. We calculate bolometric fluxes (accurate to 10--20\%) for 101 late-M and L dwarfs from SDSS, 2MASS, and WISE photometry, SDSS spectra, and BT-Settl model spectra. We find that the $z$- and $J$-band bolometric corrections for late-M and L dwarfs have a strong correlation with $z-J$ and $J-K_S$ colors respectively. The new $\chi$ values, which can be used to convert H$\alpha$ equivalent widths to activity strength, are based on spectrophotometrically calibrated SDSS spectra and the new bolometric fluxes. While the measured $\chi$ values have typical uncertainties of $\sim$20\%, ultracool dwarf models show the continuum surrounding H$\alpha$ can vary by up to an order of magnitude with changing surface gravity. Our semi-empirical $\chi$ values are one to two orders of magnitude larger than previous $\chi$ values for mid- to late-L dwarfs, indicating that the upper limits for H$\alpha$ activity strength on the coolest L dwarfs have been underestimated. 
\end{abstract}

\section{Introduction}
\label{sec:intro}
Ultracool dwarfs (here, spectral types M7-L8) include both the least massive stars and warmest brown dwarfs \citep{Burrows1997,Chabrier2000a,Dieterich2014}. Over the past fifteen years, the ultracool dwarf subfield has transformed from an effort to identify and classify hundreds of these low mass objects to the characterization of thousands of late-M and L dwarfs. As larger samples of ultracool dwarfs are being examined, it is increasingly important to measure accurate properties, including spectral type, radial velocity, bolometric fluxes, and $\chi$  values which are used to convert H$\alpha$ EW to activity strength. 

Optically defined L spectral types are anchored to the spectra of specific dwarfs identified in initial L dwarf surveys \citep[e.g.,][]{Kirkpatrick1999, Martin1999}. The use of single objects as spectroscopic templates are subject to the peculiarities of each dwarf \citep[e.g., the original L2 standard, Kelu-I, is an unresolved binary;][]{Liu2005}. In contrast, median spectra of tens to hundreds of L dwarfs per spectral subclass represent the spectral properties of the ensemble and can be used as a reliable tool to investigate spectroscopic peculiarities of individual ultracool dwarfs. We present L dwarf templates based on spectra from the Sloan Digital Sky Survey \citep[SDSS;][]{York2000} and additional late-type L dwarf spectra from the Dual Imaging Spectrograph (DIS) on the Astrophysical Research Consortium (ARC) 3.5-m telescope at Apache Point Observatory (APO). These templates, calculated using methods adopted from \citet{Bochanski2007a}, are both spectral standards and radial velocity (RV) standards that yield RVs with precisions of 7--20 km s$^{-1}$. The L dwarf templates are publicly available\footnote{\url{http://www.astro.washington.edu/users/slh/templates/ltemplates/}}.

Previously, bolometric fluxes for L dwarfs were derived using a combination of infrared spectra, model spectra, and $JHKL'M'$ photometry \citep{Leggett2001,Golimowski2004}. In the last decade, there have been dramatic improvements in model atmospheres \citep[mostly centered around the treatment of clouds and dust condensation; e.g.,][]{Allard2011} and a large increase in the number of dwarfs with mid-infrared photometry due to data from the Wide-Field Infrared Survey Explorer \citep[WISE;][]{Wright2010}. We present bolometric fluxes for ultracool dwarfs based on SDSS spectra and photometry, mid-infrared photometry from WISE, near-infrared photometry from the Two Micron All-Sky Survey \citep[2MASS;][]{Skrutskie2006}, and the BT-Settl grid of model spectra generated by a modified version of the PHOENIX atmosphere code \citep{Allard2011}.  Using these bolometric fluxes, we derive relations between bolometric corrections in the $z$ band and both $z-J$ color and spectral type and the $J$ and $K_S$ bands and $J-K_S$ color. 

Despite predictions that chromospheric activity should be rare on L dwarfs due to their cool, mostly neutral atmospheres \citep{Mohanty2002}, H$\alpha$ emission has been detected in the optical spectra of $\sim$50 early- to mid- L dwarfs \citep{Gizis2000, Kirkpatrick2000, Hall2002, Mohanty2003, Liebert2003, Schmidt2007, Reiners2008} and is likely to be present in $\sim$50\% of all L0-L5 dwarfs (though it is difficult to detect because the emission is relatively weak; S. J. Schmidt, 2014, in prep.). Interactions between the magnetic field and a mostly neutral atmosphere should be rare, but it is possible that stochastic processes \citep[such as the dust charge interactions described by][]{Helling2011b,Helling2011a,Helling2013} could be responsible for chromospheric heating. The strength of chromospheric activity, measured by the ratio of H$\alpha$ luminosity to bolometric luminosity \citep[$L_{\rm H\alpha}/L_{\rm bol}$; e.g.,][]{Hawley1996}, is an important constraint on the interaction between magnetic fields and L dwarf atmospheres. 

Activity strength declines steadily from a median of log($L_{\rm H\alpha}/L_{\rm bol}$) $\sim-4.2$ for M7 dwarfs, to values ranging from log($L_{\rm H\alpha}/L_{\rm bol}$) = $-5.8$ to $-7$ for mid-L dwarfs \citep{Schmidt2007,Reiners2008}. The precise activity strength for L dwarfs depends strongly on the conversion of the H$\alpha$ equivalent widths (EW)  to $L_{\rm H\alpha}/L_{\rm bol}$, which is frequently done using the $\chi$ value. The $\chi$ value, introduced by \citet{Walkowicz2004}, is the ratio of the continuum flux surrounding a line (in this case H$\alpha$) to the bolometric flux. $L_{\rm H\alpha}/L_{\rm bol}$ is calculated by multiplying the H$\alpha$ EW by the $\chi$ value for a given spectral type or color. The only available $\chi$ values for L dwarfs \citep{Reiners2008} are based entirely on PHOENIX DUSTY models \citep{Allard2001} that have been updated with a cloud model that better reproduces the mean properties of late-M and L dwarfs \citep[BT-Settl models;][]{Allard2011}. The continuum surrounding H$\alpha$ emission is very sensitive to metallicity, gravity, and condensation, so an entirely model based $\chi$ may not be a realistic representation of the actual value. Additionally, a model based $\chi$ for L dwarfs requires a conversion of effective temperature and spectral type. Such conversions have significant scatter in the mid- to late-L dwarf regime \citep{Golimowski2004,Stephens2009}. We calculate new semi-empirical $\chi$ values using a combination of our new bolometric fluxes and spectrophotometrically calibrated SDSS spectra, avoiding the use of the model continuum near H$\alpha$ and removing the need for $T_{\rm eff}$ conversion. We present relations for $\chi$ as a function of $i-z$ color, $i-J$ color, and spectral type. 

In Section~\ref{sec:data}, we describe the selection of the spectroscopic and photometric data and examine the BT-Settl model grid. In Section~\ref{sec:templ} we discuss the calculation and fidelity of the templates, and in Section~\ref{sec:fbol} we describe the calculation of bolometric fluxes. Section~\ref{sec:chi} includes our description of the calculation of $\chi$ values and the comparison of those values to previous work. 

\section{Data and Models}
\label{sec:data}
Our calculation of spectroscopic templates, bolometric fluxes, and $\chi$ values are based on spectroscopic and photometric data from multiple sources in addition to the BT-Settl model grid. The templates include spectra both from SDSS and the ARC 3.5-m telescope. The bolometric fluxes and $\chi$ values are calculated based on WISE, 2MASS, and SDSS photometry combined with SDSS spectroscopy and BT-Settl model spectra. The selection and properties of the data and the models are described below. 

\subsection{Spectroscopic Data for Late-M and L Dwarfs}
\begin{figure}
\includegraphics[width=\linewidth]{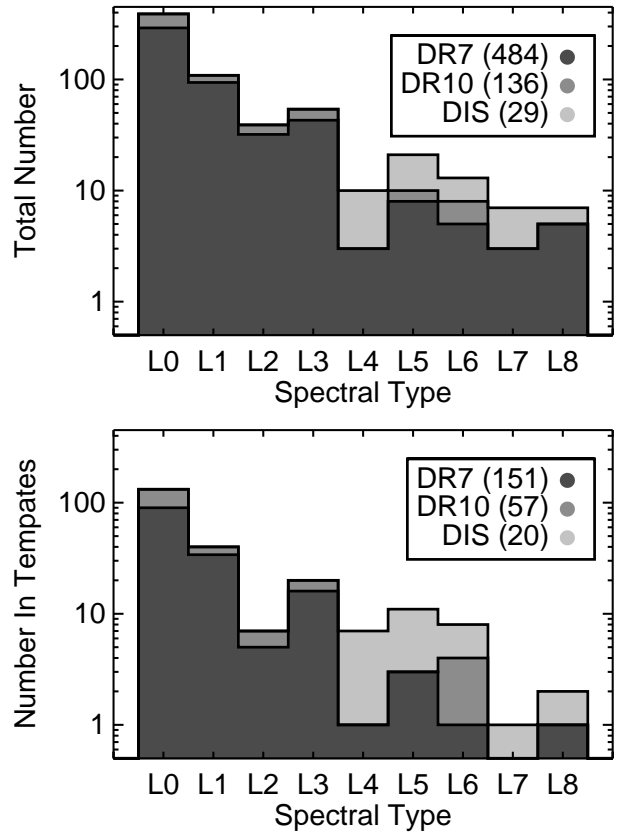} 
\caption{Spectral type distribution of the total number of spectra from DR7, DR10, and DIS available for use in the templates (top) and the number of spectra actually used in the final templates (as described in Section~\ref{sec:rvl}; bottom). The large number of early-L dwarfs in DR7 is due primarily to their intrinsically larger luminosities compared to later-L dwarfs.} \label{fig:st}
\end{figure}

The spectra available for inclusion in the templates are from three different sources: 1) The majority of the spectra are from the sample of 484 L0-L8 dwarfs selected by \citet[][hereafter S10]{Schmidt2010} from the Seventh Data Release of SDSS \citep[DR7;][]{Abazajian2009}; 2) 136 additional L0-L8 dwarf spectra from \citet{Schmidt2012Phd} that are part of the SDSS-III Baryon Oscillation Sky Survey \citep[BOSS;][]{Dawson2013}; and 3) observations of 29 L4-L8 dwarfs using DIS on the ARC 3.5-m telescope. The spectral type distribution of the 649 L dwarf spectra from these three sources is shown in Figure~\ref{fig:st}.

The DR7 L dwarf sample is described in more detail by S10, but we summarize as follows. The DR7 L dwarfs were selected first by a single color cut ($i-z>1.4$) designed to include the bluest outliers in the $i-z$ distribution of L0 dwarfs \citep[observed to have a median $i-z$=1.84 and dispersion of 0.06;][]{West2008}. The color cut resulted in over 13,000 spectra of red objects in SDSS. Each spectrum was spectral typed using the Hammer spectral typing software \citep{West2004,Covey2007} and the types were visually inspected, resulting in a final sample of 484 L dwarfs. 

A total of 10,000 BOSS fibers have been awarded for our BOSS Ultracool Dwarf (BUD) ancillary program designed to increase the number of late-M and L dwarfs with SDSS spectra \citep[][Schmidt2012phd]{Dawson2013}. Just over half (5007) of the candidate late-M and L dwarfs were observed as part of the tenth data release \citep[DR10;][]{Ahn2014} of SDSS. Approximately half of those were ultracool dwarfs, and when combined with ultracool dwarfs targeted by other portions of the survey, the total sample from DR10 is 2540 M7-M9 dwarf and 136 L dwarfs. The selection and spectral typing of these ultracool dwarfs is similar to that of the DR7 L dwarfs and is described in detail by \citet{Schmidt2012Phd}.

While there are 620 L dwarfs with spectra in DR7 and DR10, the majority have L0-L3 spectral types, with only 23 L4-L8 dwarfs (shown in Figure~\ref{fig:st}). This is due primarily to the faintness of mid- to late-L dwarfs (the absolute $i$ magnitude falls from $M_i \sim 15.5$ for L0 dwarfs to $M_i \sim 18.5$ for L5 dwarfs and $M_i \sim 20.5$ for L8 dwarfs; S10). To supplement the small numbers of mid- to late- L dwarfs, we conducted additional observations of 29 L4-L8 dwarfs using DIS on the ARC 3.5-m telescope. Our supplemental list of L4-L8 dwarfs was obtained from Dwarf Archives\footnote{\url{http://DwarfArchives.org}}; the target list included the brightest objects in each spectral type bin. We chose to exclude the L9 spectral class due to their faint magnitudes and the lack of L9 dwarfs in original SDSS spectroscopy. The details of the additional L4-L8 dwarfs and their observations are given in Table~\ref{tab:obs}.

\tabletypesize{\tiny}
\begin{deluxetable*}{llllllllllll} \tablewidth{0pt} 
\tablecaption{Observed L dwarfs \label{tab:obs}}
\tablehead{\colhead{Designation\tablenotemark{a}} & \colhead{R.A.}  & \colhead{decl.}   & \colhead{ST}  & \colhead{Ref.}  & \colhead{$J$} & \colhead{$K_S$}   & \colhead{Obs. Date}  & \colhead{N$_{\textrm{obs}}$}  & \colhead{$T_{\textrm{exp}}$}  & \colhead{$T_{\textrm{obs}}$} & \colhead{Weather} }
\startdata
SDSS J000112.18+153535.5   	&	00 01 12.1	&	+15 35 35.5	&	L4	&	1	&	15.52	$\pm$	0.06	&	13.71	$\pm$	0.04	&	2011 Jan 5	&	4	&	900	&	3600	&	mostly clear	\\
2MASS J01311838+3801554    	&	01 31 18.3	&	+38 01 55.4	&	L4	&	2	&	14.68	$\pm$	0.03	&	13.05	$\pm$	0.03	&	2011 Jan 5	&	3	&	750	&	2250	&	mostly clear	\\
2MASS J05002100+0330501   	&	05 00 21.0	&	+03 30 50.1	&	L4	&	3	&	13.67	$\pm$	0.02	&	12.06	$\pm$	0.02	&	2010 Oct 31	&	3	&	300	&	900	&	clear	\\
2MASS J05012406$-$0010452    	&	05 01 24.0	&	$-$00 10 45.2	&	L4	&	3,4	&	14.98	$\pm$	0.04	&	12.96	$\pm$	0.03	&	2011 Jan 5	&	3	&	900	&	2700	&	mostly clear	\\
DENIS-P J153941.96$-$052042.4	&	15 39 41.9	&	$-$05 20 42.8	&	L4	&	5,6	&	13.92	$\pm$	0.03	&	12.58	$\pm$	0.03	&	2011 Jun 29	&	3	&	600	&	1800	&	partly cloudy	\\
2MASS J18212815+1414010   	&	18 21 28.1	&	+14 14 01.0	&	L4	&	7	&	13.43	$\pm$	0.02	&	11.65	$\pm$	0.02	&	2011 Jun 9	&	3	&	300	&	900	&	clear	\\
2MASSW J2224438$-$015852     	&	22 24 43.8	&	$-$01 58 52.1	&	L4	&	8	&	14.07	$\pm$	0.03	&	12.02	$\pm$	0.02	&	2011 Jun 9	&	2	&	400	&	800	&	clear	\\
2MASS J03552337+1133437   	&	03 55 23.3	&	+11 33 43.7	&	L5	&	3,4	&	14.05	$\pm$	0.02	&	11.53	$\pm$	0.02	&	2011 Jan 12	&	3	&	900	&	2700	&	partly cloudy	\\
SDSSp J053951.99$-$005902.0  	&	05 39 52.0	&	$-$00 59 01.9	&	L5	&	9	&	14.03	$\pm$	0.03	&	12.53	$\pm$	0.02	&	2010 Oct 31	&	4	&	300	&	1200	&	clear	\\
2MASSI J0652307+471034      	&	06 52 30.7	&	+47 10 34.8	&	L5	&	10	&	13.51	$\pm$	0.02	&	11.69	$\pm$	0.02	&	2010 Oct 31	&	3	&	250	&	750	&	clear	\\
SDSS J080531.84+481233.0   	&	08 05 31.8	&	+48 12 33.1	&	L5	&	11	&	14.73	$\pm$	0.03	&	13.44	$\pm$	0.04	&	2011 Jan 5	&	3	&	800	&	2400	&	mostly clear	\\
2MASSI J0835425$-$081923       	&	08 35 42.5	&	$-$08 19 23.7	&	L5	&	10	&	13.17	$\pm$	0.02	&	11.14	$\pm$	0.02	&	2010 Oct 31	&	3	&	200	&	600	&	clear	\\
2MASS J09054654+5623117   	&	09 05 46.5	&	+56 23 11.7	&	L5	&	3	&	15.40	$\pm$	0.05	&	13.73	$\pm$	0.04	&	2010 Oct 31	&	3	&	900	&	2700	&	clear	\\
2MASSW J1507476$-$162738     	&	15 07 47.6	&	$-$16 27 38.6	&	L5	&	8,12	&	12.83	$\pm$	0.03	&	11.31	$\pm$	0.02	&	2010 Jul 11	&	1	&	1800	&	1800	&	clear	\\
SDSS J171714.10+652622.2   	&	17 17 14.0	&	+65 26 22.1	&	L5	&	11	&	14.95	$\pm$	0.04	&	13.18	$\pm$	0.03	&	2010 Jul 7	&	4	&	600	&	2400	&	mostly clear	\\
2MASS J17461199+5034036    	&	17 46 11.9	&	+50 34 03.6	&	L5	&	3	&	15.10	$\pm$	0.06	&	13.53	$\pm$	0.04	&	2010 Jul 7	&	2	&	600	&	1200	&	mostly clear	\\
2MASS J17502484$-$0016151    	&	17 50 24.8	&	$-$00 16 15.1	&	L5	&	13	&	13.29	$\pm$	0.02	&	11.85	$\pm$	0.02	&	2010 Jul 7	&	2	&	600	&	1200	&	mostly clear	\\
2MASS J21373742+0808463   	&	21 37 37.4	&	+08 08 46.3	&	L5	&	3	&	14.77	$\pm$	0.03	&	13.02	$\pm$	0.03	&	2011 Jun 9	&	3	&	800	&	2400	&	clear	\\
2MASS J01443536$-$0716142    	&	01 44 35.3	&	$-$07 16 14.2	&	L6	&	14	&	14.19	$\pm$	0.02	&	12.27	$\pm$	0.02	&	2010 Oct 31	&	3	&	500	&	1500	&	clear	\\
SDSS J065405.63+652805.4   	&	06 54 05.6	&	+65 28 05.1	&	L6	&	15	&	16.14	$\pm$	0.09	&	14.60	$\pm$	0.08	&	2011 Jan 12	&	6	&	1200	&	7200	&	partly cloudy	\\
2MASS J20025073$-$0521524    	&	20 02 50.7	&	$-$05 21 52.4	&	L6	&	2	&	15.32	$\pm$	0.05	&	13.42	$\pm$	0.04	&	2011 Jun 29	&	3	&	1200	&	3600	&	partly cloudy	\\
2MASS J21481628+4003593   	&	21 48 16.3	&	+40 03 59.4	&	L6	&	7	&	14.15	$\pm$	0.03	&	11.77	$\pm$	0.02	&	2011 Jun 29	&	3	&	600	&	1800	&	partly cloudy	\\
DENIS-P J225210.73$-$173013.4	&	22 52 10.7	&	$-$17 30 13.4	&	L6	&	6	&	14.31	$\pm$	0.03	&	12.90	$\pm$	0.02	&	2011 Jun 29	&	3	&	720	&	2160	&	partly cloudy	\\
SDSSp J042348.57$-$041403.5  	&	04 23 48.5	&	$-$04 14 03.5	&	L7	&	10,16	&	14.47	$\pm$	0.03	&	12.93	$\pm$	0.03	&	2010 Oct 31	&	4	&	600	&	2400	&	clear	\\
2MASSI J0439010$-$235308       	&	04 39 01.0	&	$-$23 53 08.3	&	L7	&	10	&	14.41	$\pm$	0.03	&	12.82	$\pm$	0.02	&	2010 Oct 31	&	4	&	500	&	2000	&	clear	\\
2MASS J09153413+0422045    	&	09 15 34.1	&	+04 22 04.5	&	L7	&	3	&	14.55	$\pm$	0.03	&	13.01	$\pm$	0.04	&	2011 Jan 5	&	3	&	720	&	2160	&	mostly clear	\\
2MASSI J1526140+204341      	&	15 26 14.0	&	+20 43 41.4	&	L7	&	8	&	15.59	$\pm$	0.05	&	13.92	$\pm$	0.05	&	2011 Jun 9	&	5	&	900	&	4500	&	clear	\\
DENIS-P J0205.4$-$1159              	&	02 05 29.4	&	$-$11 59 29.6	&	L8	&	17,18	&	14.59	$\pm$	0.03	&	13.00	$\pm$	0.03	&	2010 Oct 31	&	4	&	800	&	3200	&	clear	\\
2MASS J02572581$-$3105523    	&	02 57 25.8	&	$-$31 05 52.3	&	L8	&	5	&	14.67	$\pm$	0.04	&	12.88	$\pm$	0.03	&	2011 Jan 5	&	4	&	900	&	2600	&	mostly clear	
\enddata
\tablenotetext{a}{Based on discovery designation from DwarfArchives.org.}
\tablerefs{(1) \citet{Knapp2004}; (2) \citet{Cruz2007}; (3) \citet{Reid2008}; (4) \citet{Cruz2009}; (5) \citet{Kirkpatrick2008}; (6) \citet{Kendall2004}; (7) \citet{Looper2008a}; (8) \citet{Kirkpatrick2000}; (9) \citet{Fan2000}; (10) \citet{Cruz2003}; (11) \citet{Hawley2002}; (12) \citet{Reid2000}; (13) \citet{Kendall2007}; (14) \citet{Liebert2003}; (15) \citet{Chiu2006}; (16) \citet{Geballe2002}; (17) \citet{Kirkpatrick1999}; (18) \citet{Delfosse1997}}
\end{deluxetable*}
\tabletypesize{\footnotesize}

The DIS spectra were obtained with the R300 grating, a wavecenter of 7500\AA, and a $1\arcsec.5$ slit, resulting in R$\sim$1000 spectra covering from 5200\AA~to 10000\AA. Our total exposure times (given in Table~\ref{tab:obs}) were calculated to achieve a rough S/N$\sim$10. Each object was observed in 3-5 exposures to effectively remove cosmic rays.  The data were reduced using standard IRAF\footnote{IRAF is distributed by the National Optical Astronomy Observatories, which are operated by the Association of Universities for Research in Astronomy, Inc., under cooperative agreement with the National Science Foundation.} procedures. We used a HeNeAr lamp spectrum for wavelength calibration and flux-calibrated each spectrum using spectrophotometric standards. 

The individual spectra for each object were corrected to a heliocentric rest frame and combined using a custom IDL code. While each dwarf already had a spectral type assigned, some were based on infrared spectra. We therefore re-typed each spectrum using the Hammer spectral typing software \citep{West2004,Covey2007} and visually inspected each type. The spectral types we use are given in Table~\ref{tab:obs} and only differ from previously published types by $\pm$1 subtype. 

In addition to using spectra for the templates, we also incorporated the SDSS spectra in our estimation of bolometric fluxes for late-M and L dwarfs. Only a fraction of available spectra were used due to strong cuts on the quality of photometry for each object included (as described in Section~\ref{sec:fbol}). In addition to a subset of the SDSS L dwarf spectra described above, we also used 65 M7-M9 dwarf spectra from the SDSS DR7 M dwarf catalog \citep{West2011} and eleven spectra from the BUD sample  \citep{Schmidt2012Phd}. The selection and spectral typing of those spectra was similar to that of the SDSS and BUD L dwarfs and is described in detail in \citet{West2011} and \citet{Schmidt2012Phd} respectively. 

\subsection{Photometric Data}
\label{sec:photdat}
To constrain the bolometric fluxes of late-M and L dwarfs (as described in Section~\ref{sec:fbol}), we supplemented SDSS spectra with SDSS, 2MASS, and WISE photometry.  The selection of the photometry is described in more detail in \citet{Schmidt2012Phd}, but we summarize here. SDSS photometry was obtained from the DR7 database via a coordinate cross-match\footnote{Using the online SDSS CrossID tool at \url{http://cas.sdss.org/dr7/en/tools/crossid/crossid.asp}} between the photometric and spectroscopic databases of SDSS. We exclude objects with the SATURATED, BAD\_COUNTS\_ERROR, INTERP\_CENTER, PSF\_FLUX\_INTERP, or NODEBLEND flags set in either the $i$ or the $z$ band. 

We obtained photometry from 2MASS based on a match to the closest source within 5\arcsec~of the SDSS coordinates. The search radius is a balance between including objects with relatively high proper motions (SDSS and 2MASS images are 0--10 yrs apart, depending on the field) and excluding false matches where the source was not detected. The faintest 5\% of the SDSS sources did not match to 2MASS due to the shallower magnitude limits of 2MASS. For good quality photometry, we required the rd\_flg = 2, cc\_flg=0, and ph\_qual=ABCD flags for each magnitude. The WISE photometry for each dwarf was selected based on a match to the closest source within 5\arcsec~of the SDSS coordinates. We included only WISE matches with ext\_flg $\ne$ 2 and nb = 1 to exclude sources there were blended due do their proximity to another source.  We also required cc\_flg = 0, ph\_qual=ABC, and a real (not null) uncertainty to ensure good quality photometry. With these cuts in place, 3\% of the original SDSS sources were not matched in WISE, primarily due proximity to other sources (leading either to a blending flag or poor quality photometry). Additional information on the matching, including checks on source confusion, can be found in 
Schmidt et al. (2014, in prep.).

In addition to the flag cuts described above, we excluded photometry with uncertainties greater than those listed in Table~\ref{tab:unc}. Each uncertainty cut was chosen by fitting a gaussian to the uncertainty distribution then excluding photometry with uncertainties greater than the mean of the gaussian plus twice the standard deviation. For the $W3$ band, the mean fell above 0.5 so 0.15 was selected. Additional cuts were performed on the WISE $W1$ and $W2$ bands due to a systematic faint magnitude bias described in Section~6.3 of the All-Sky Release Supplement\footnote{\url{http://wise2.ipac.caltech.edu/docs/release/allsky/expsup/sec6\_3c.html}}. We excluded magnitudes of $W1 > 14$ and $W2 >$ 14.5, corresponding to the points where the bias is comparable to the average uncertainties. The limiting magnitude of the WISE $W4$ band is too bright to include the majority of our sources, so we excluded it from this analysis. 

\renewcommand{\arraystretch}{0.8}
\begin{deluxetable}{lll} \tablewidth{0pt} \tabletypesize{\scriptsize}
\tablecaption{Uncertainty Cuts \label{tab:unc} }
\tablehead{ \colhead{Band}  & \colhead{$\sigma_m$} &  \colhead{Magnitude}\\  \colhead{ }  & \colhead{Limit} &  \colhead{Limit}  } 
\startdata
$r$ & 0.10 & \nodata \\
$i$ & 0.03 & \nodata \\
$z$ & 0.04 & \nodata \\
$J$ & 0.17 & \nodata \\
$H$ & 0.22 & \nodata \\
$K_S$ & 0.26 & \nodata \\
$W1$ & 0.05 & 14.5 \\
$W2$ & 0.13 & 14 \\
$W3$ & 0.15 & \nodata 
\enddata
\end{deluxetable}
\renewcommand{\arraystretch}{1}

\subsection{Understanding the Model Grid}
\label{sec:describe_mod}
The atmospheres of late-M an L dwarfs ($1200 \lesssim T_{\rm eff} \gtrsim 2400$~K) are complicated by the condensation of dust grains and the formation of a cloud deck \citep[e..g,][]{Tsuji1996a,Ackerman2001}. The efficiency of dust formation and the thickness of the cloud deck have a variety of effects on the stellar atmosphere (e.g., the disappearance of TiO bands in the optical and the strength of collisional H$_{\alpha}$ absorption in the infrared), and those features are observed to vary not only with $T_{\rm eff}$ but also with age, log(g), and [M/H]. There are many different approaches to modeling these effects \citep[see][for an overview]{Helling2008b}, but the most common is to tune the thickness of the clouds with a single parameter \citep[e.g., $f_{sed}$;][]{Ackerman2001}. 

The BT-Settl grid of model spectra \citep{Allard2011} is generated by adding a radiative hydrodynamic cloud model to the PHOENIX code. Particles are allowed to heat on a lower level and convect upward, then sediment and settle out. Instead of using a parameter to tune the cloud thickness, the clouds evolve depending on the gravity, temperature, and metallicity. In addition to stronger physical motivations, the BT-Settl cloud model also reproduces the mean $J-K_S$ colors of late-M and L dwarfs better than other model grids. 

\begin{figure}
\begin{center}
\includegraphics[width=0.9\linewidth]{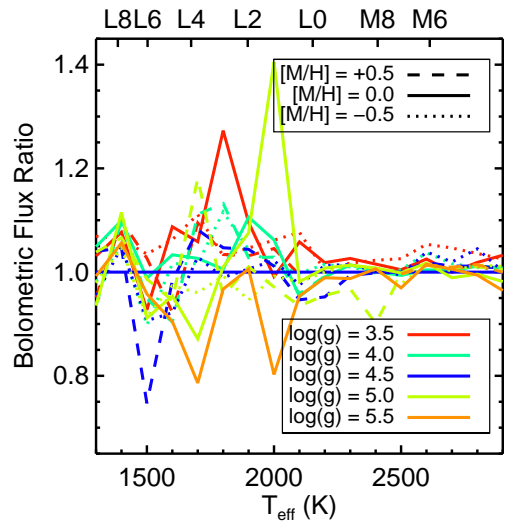} 
\caption{The ratio of bolometric flux for each model compared to the [M/H] = 0.0, log(g) = 4.5 model of the same $T_{\rm eff}$. Each value of log(g) is indicated by a different color, while [M/H] is indicated by a different linestyle. The variations of bolometric flux with respect to [M/H] or log(g) are smaller than the typical uncertainties in bolometric flux (10--20\%).} \label{fig:ratio}
\end{center}
\end{figure}

Rather than using a precise fit between the model and data to determine the properties of each object, we aim only to use the overall spectral energy distribution that best matches each object to determine its bolometric flux ($F_{bol}$). To understand the variation of $F_{bol}$ with changes in the model parameters, we examine $F_{bol}$ for each $T_{\rm eff}$ as we vary [M/H] and log(g). We investigate the BT-Settl model grid spanning $T_{\rm eff} = 1300$--$3100$~K (in steps of 100~K), log(g) = 3.5--5.5 (in steps of 0.5 dex), and [M/H] = $-0.5$, $0.0$, and $+0.5$. For each model, we integrated over the full spectrum to determine $F_{bol}$. Figure~\ref{fig:ratio} shows the ratio of $F_{bol}$ for each model compared to the [M/H] = 0.0 and log(g) = 4.5 model for each $T_{\rm eff}$. 

Variations in the bolometric flux with log(g) and [M/H] are on, on average, at the $\sim10$\% level (though in four cases are $>$20\%) and not systematic. Even without a perfect match to log(g) and [M/H], we can determine $F_{bol}$ with reasonable accuracy and precision. In Section~\ref{sec:fbol}, we describe the fits between the data and model spectra to calculate the bolometric fluxes for late-M and L dwarfs. We fit both $T_{\rm eff}$ and log(g) and adopt uncertainties that include the full range of log(g) for each object. We do not fit [M/H], in part due to the lack of a complete [M/H] grid available for all $T_{eff}$ and log(g), and in part because the variations in bolometric flux due to [M/H] are on a similar scale to those for log(g). 

\section{L Dwarf Spectroscopic Templates}
\label{sec:templ}
The templates were calculated using a method adapted from \citet{Bochanski2007a}. The template spectra have higher S/N than any individual spectra due to the coaddition of flux, but they also can have increased wavelength resolution. SDSS spectra are typically interpolated onto the same wavelength grid, but correcting each one to zero radial velocity shifts wavelength with respected to pixel and allows a finer wavelength sampling than the original spectra. The final templates are then suitable both as spectral type and radial velocity standards.

\subsection{Radial Velocities for L dwarfs}
\label{sec:rvl}
To ensure that the final templates are accurate RV standards, we measured new velocities for each L dwarf included in the templates (despite some L dwarfs having previously measured velocities by S10). The SDSS spectra (both DR7 and DR10) are wavelength-calibrated to a vacuum scale and corrected to a heliocentric frame of rest. For consistency with the DIS spectra, which were wavelength-calibrated to an air scale, we converted the SDSS spectra to an air scale before measuring velocities.

We measured radial velocities for the L dwarfs using as many as seven atomic lines (H$\alpha$, Rb I $\lambda$8900\AA~and $\lambda$7948\AA, Na I $\lambda$8183\AA~and $\lambda$8194\AA, and Cs I $\lambda$8521\AA~and $\lambda$8943\AA). Each line was first measured and those with equivalent widths $<1$~\AA~were rejected as non-detections. Then each detected line was fit by a gaussian function, with the mean of the gaussian indicating the stellar velocity as measured by that line. We then calculated the mean and standard deviation of the measured velocities from all the detected lines. If the standard deviation was greater than 25~km~s$^{-1}$, we iteratively rejected the farthest line from the mean until the standard deviation was less than 25~km~s$^{-1}$. We required at least two measurements with a standard deviation below 25~km~s$^{-1}$ to use each spectrum. The templates were calculated from these 228 L dwarfs with good velocities (their spectral type distribution is shown in Figure~\ref{fig:st}). 

\subsection{Template Calculation}
We first corrected each spectrum with a measured radial velocity to zero radial velocity, then normalized each spectrum between 8300 and 8400\AA~(this region is as close to continuum as possible for the early-L dwarfs, and represents a balance between sufficient flux and avoiding calibration issues for late-L dwarfs) and interpolated the fluxes and uncertainties onto a single wavelength array. Due to variable sample size and quality in each spectral type bin, we chose different velocity resolutions for each template of a given spectral sub-type (values are listed in Table~\ref{tab:rv}). We chose to generate a wavelength array with logarithmic spacing, resulting in wavelength pixels that have constant velocity spacing. The constant velocity spacing allows cross-correlation over a large region of the spectrum without losing velocity information to interpolation. 

\renewcommand{\arraystretch}{0.8}
\begin{deluxetable*}{lllllllllll} \tablewidth{0pt} \tabletypesize{\scriptsize}
\tablecaption{Properties of the Templates as RV Standards \label{tab:rv} }
\tablehead{ \colhead{Spectral}  & \colhead{Resolution}  & \colhead{Resolution} &  \multicolumn{8}{c}{RV (km s$^{-1}$)}\\  
\colhead{Type}  & \colhead{per pixel} &  \colhead{per pixel}  &  \colhead{H$\alpha$} &  \colhead{Rb I}  &  \colhead{Rb I} &  \colhead{Na I}  &  \colhead{Na I} &  \colhead{Cs I}  &  \colhead{Cs I} &  \colhead{mean and} \\
\colhead{}  & \colhead{(km s$^{-1}$)} &  \colhead{$\lambda/\Delta\lambda$ }  &  \colhead{} &  \colhead{$\lambda$8900\AA}  &  \colhead{$\lambda$7948\AA} &  \colhead{$\lambda$8183\AA}  &  \colhead{$\lambda$8194\AA} &  \colhead{$\lambda$8521\AA}  &  \colhead{$\lambda$8943\AA} &  \colhead{dispersion} 
 } 
\startdata
L0 & 30 &  10000 & $-$35.0 & $-$30.3 & \nodata & $-$3.5 & \phs10.9 & $-$47.2 & \nodata & $-$21.0 $\pm$ 23.9\\
L1 & 30 &  10000 & \phs2.0 & $-$8.4 & \phs2.4 & \phs4.0 & \phs13.1 & $-$10.6 & $-$18.0 & $-$2.2 $\pm$ 10.5\\
L2 & 45 &  6700 & $-$22.6 & \phs0.5 & \phs2.1 & $-$14.7 & \phs7.9 & $-$1.6 & \phs36.2 & \phs1.1 $\pm$ 18.7\\
L3 & 45 &  6700 & \nodata & $-$18.9 & $-$12.7 & \phs3.8 & $-$2.8 & \phs1.1 & $-$55.0 & $-$14.1 $\pm$ 21.8\\
L4 & 45 &  6700 &  \nodata & \nodata & \phs11.2 & \nodata & $-$43.1 & $-$18.0 & \phs1.2 & $-$12.2 $\pm$ 23.9\\
L5 & 45 &  6700 & \nodata & \nodata & \phs5.6 & \nodata & \phs3.7 & \phs0.8 & \phs2.5 & \phs3.1 $\pm$ 2.1\\
L6 & 60 &  5000 & \nodata & \nodata & $-$18.0 & $-$8.5 & \phs4.9 & \phs7.3 & $-$12.4 & $-$5.3 $\pm$ 11.0\\
L7 & 60 &  5000 & \nodata & \nodata & \phs0.0 & \phs34.0 & \nodata & \phs1.3 & $-$25.9 & \phs2.4 $\pm$ 24.5\\
L8 & 60 & 5000 & \nodata & \nodata & \nodata & \phs12.4 & \nodata & \phs13.7 & $-$8.1 & \phs6.0 $\pm$ 12.2
\enddata
\end{deluxetable*}
\renewcommand{\arraystretch}{1}

\begin{figure*} 
\includegraphics[width=0.9\linewidth]{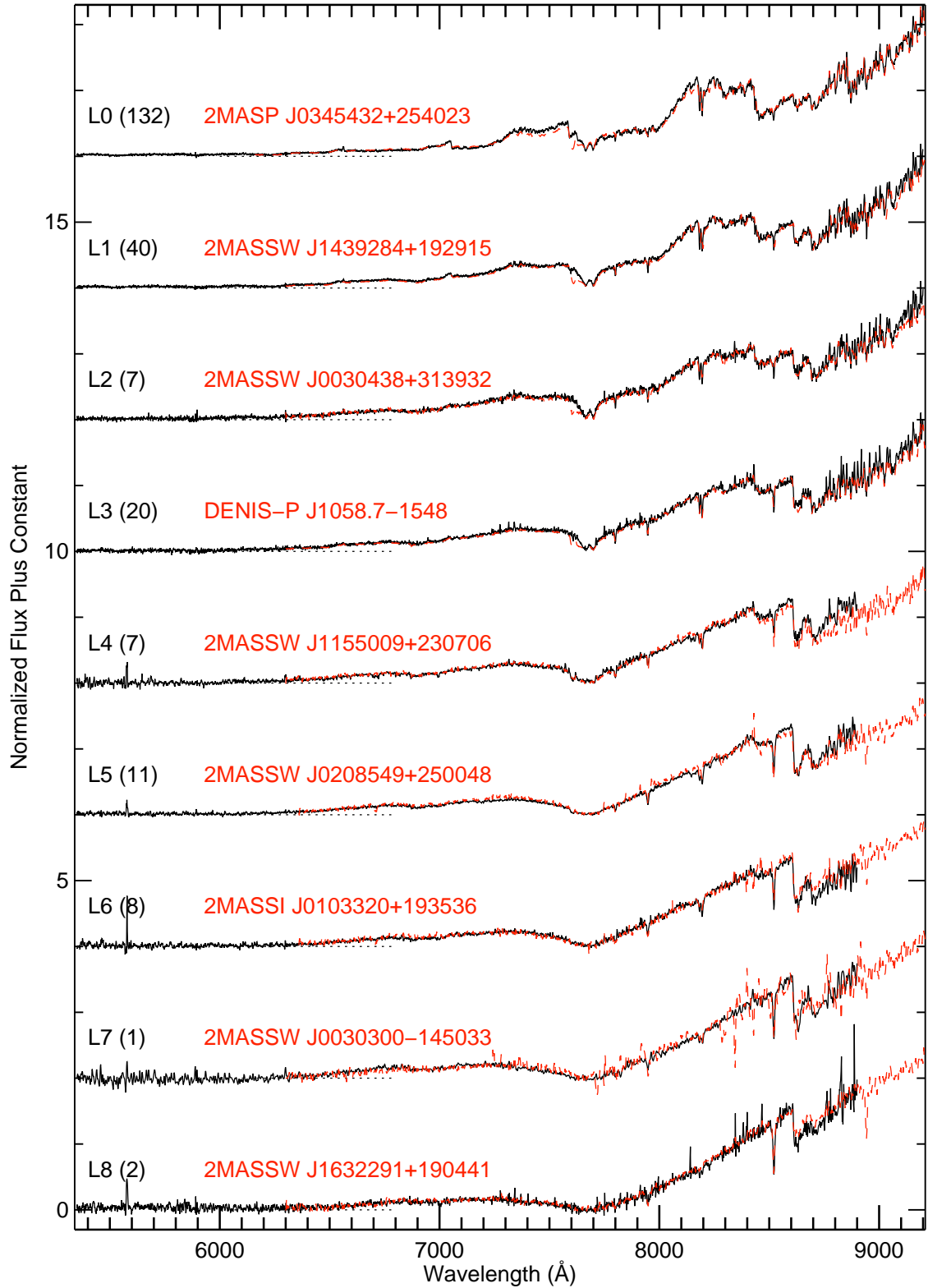} 
\caption{Spectroscopic templates for L0 to L8 dwarfs (black sold lines) compared to spectral standards (red dashed lines) from \citet{Kirkpatrick1999} and \citet{Kirkpatrick2000}. Each template is offset by a constant for clarity (dotted lines) and labeled with its spectral type and the number of spectra used and each spectral standard is labeled with its 2MASS designation. The L4-L8 templates do not extend past 8900\AA, as discussed in Section~\ref{sec:stand}.}\label{fig:templ}
\end{figure*}

The templates are the median of the normalized flux from all the spectra of each type, shown in Figure~\ref{fig:templ}. In the downloadable version of the templates\footnote{\url{http://www.astro.washington.edu/users/slh/templates/ltemplates/}}, we also include the mean and standard deviation of the normalized flux in addition to an uncertainty array calculated from the standard deviation and the uncertainties on each spectrum added in quadrature. The number of spectra included in each template is shown in Figure~\ref{fig:st} and given in Figure~\ref{fig:templ}. We note that the L7 and L8 templates are based on one and two L dwarfs, respectively. While these templates do not have the advantage of representing the mean properties of their types, they are still suitable as high S/N radial velocity standards.

\subsection{Templates as Spectral Standards}
\label{sec:stand}
To confirm that our templates are accurate spectroscopic standards for the L spectral sequence, we compared them to the spectroscopic standards from \citet{Kirkpatrick1999} and \citet{Kirkpatrick2000}\footnote{Data were obtained from the M+L LRIS spectroscopic archive found at \url{http://www.stsci.edu/$\sim$inr/lrism.html}}, shown in Figure~\ref{fig:templ}. The general agreement of both spectral slope and specific features is good. The most notable difference is a telluric feature found in the Kirkpatrick et al. spectra (and the L4 template) at 7594-7685\AA~that has been removed from SDSS spectra. The spectral slopes for L4-L8 dwarfs did not match the Kirkpatrick et al. standards at $\lambda>$8900\AA, so those wavelengths are not included in the final templates. In the L0 and L2 types, there is a mismatch between the template the Kirkpatrick et al. standard at the 7300-7500\AA~VO band, indicating that the standards may be mildly low gravity objects (see Section~\ref{sec:jkl}). Due to the combination of reduced noise and lack of spectroscopic peculiarities, the L0--L6 templates are superior to individual spectra for assigning optical types. 

\begin{figure*} 
\includegraphics[width=0.95\linewidth]{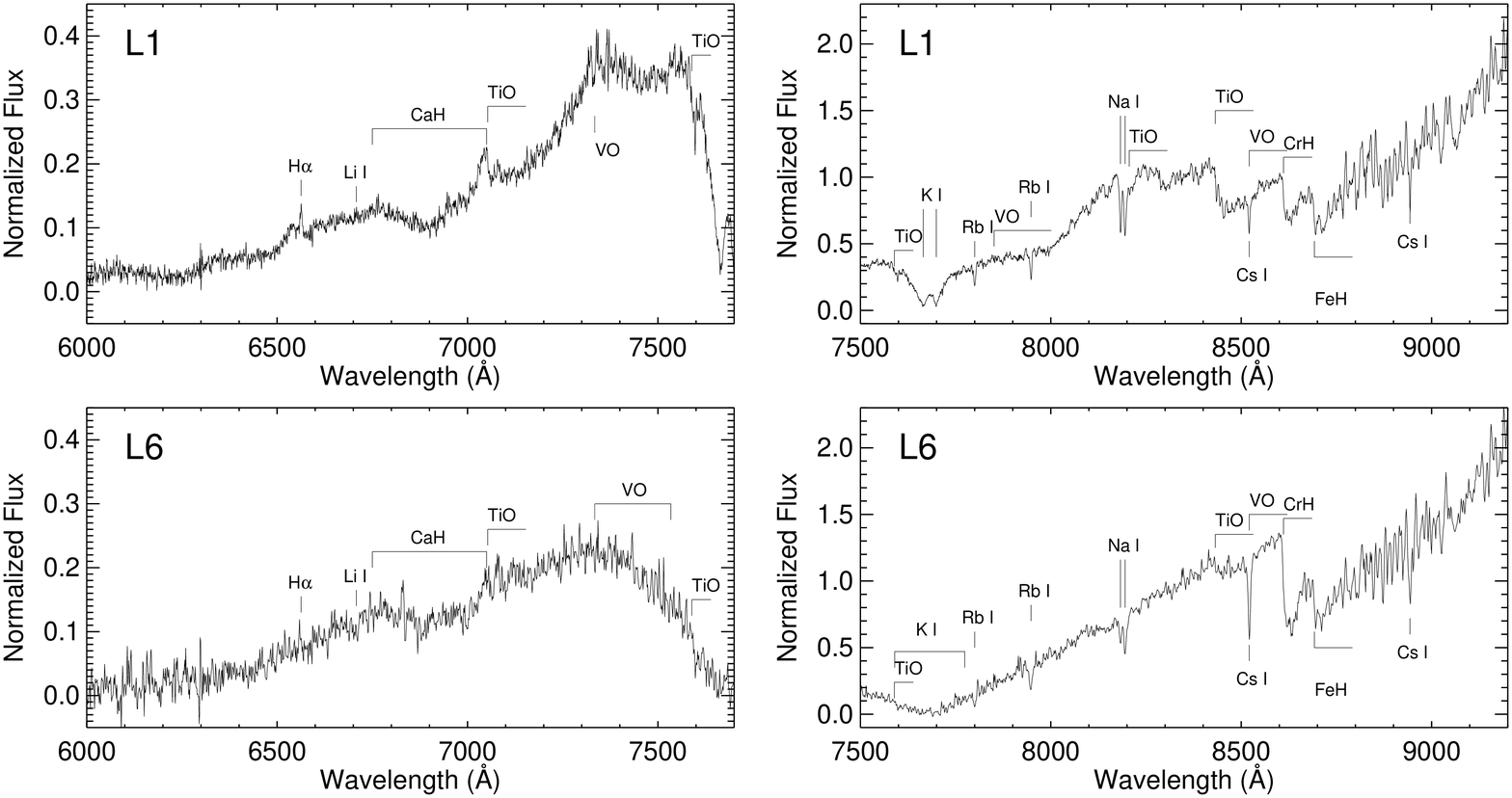} 
\caption{Template spectra of an L1 and L6 dwarf with atomic and molecular features labeled.}\label{fig:twoL}
\end{figure*}

Figure~\ref{fig:twoL} shows the L1 and L6 template spectra in greater detail with the lines and molecular band identified in \citet{Kirkpatrick1999} labeled. While Li I and H$\alpha$ are labeled on the spectra of both dwarfs, the Li I line is not present in either template and the H$\alpha$ is not present in the L6 template. H$\alpha$ is prominent in the L1 template and is visible in all L0-L3 templates. The other prominent atomic lines include the K I doublet (dramatically pressure-broadened in the L6 spectrum), Rb I, Cs I, and the Na I doublet and prominent molecular bands include CaH, TiO, VO, CrH, and FeH. 

\subsection{RV Fidelity and Uncertainties}
We examined the accuracy of the wavelength solution for each template by measuring their velocities from the same seven atomic lines used to measure velocities of individual spectra. Again, we rejected lines with equivalent widths $<1$~\AA~and iteratively rejected outliers. The resulting velocity shifts (for each line and a mean value for each template) are given in Table~\ref{tab:rv}. The mean velocities range from $-$21 to 6~km~s$^{-1}$ and are each consistent with zero within the uncertainties. The dispersion of the velocities measured by the atomic lines ranges from 10~km~s$^{-1}$ to 20~km~s$^{-1}$, indicating an average precision of $\sim$15~km~s$^{-1}$.

\begin{figure*} 
\includegraphics[width=0.95\linewidth]{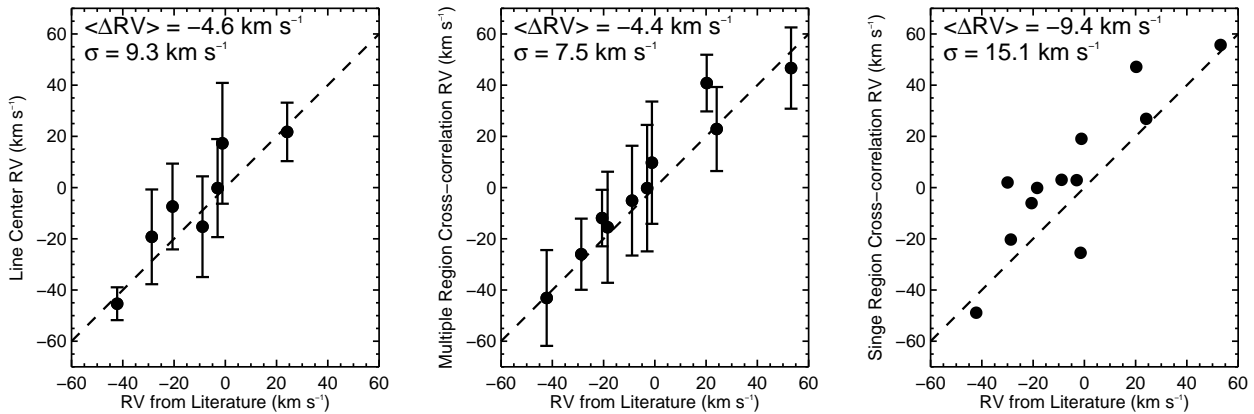} 
\caption{L dwarf RV measured using line centers (left panel), cross-correlation with templates using thirteen 200~\AA~wide regions from 6400 to 9200~\AA (middle panel), and cross-correlation over the 7000--9000\AA~range (right panel) compared to published RV measurements. The mean ($\langle\Delta$RV$\rangle$) and standard deviation ($\sigma$) of the difference between the two measurements is given in the top left corner of each panel.}\label{fig:knownv}
\end{figure*}

We also tested the accuracy and precision of the templates as velocity standards through cross-correlation with 12 L dwarfs in the SDSS database with velocities published in previous work, primarily based on high resolution spectroscopy \citep{Bailer-Jones2004,Blake2010,Schmidt2010a,Seifahrt2010}. First, we measured each velocity using the atomic line method; those results are shown in the left panel of Figure~\ref{fig:knownv}. The resulting velocities, measured for seven of the 12 L dwarfs, have a mean RV difference of $\Delta$RV=$-4.6$~km~s$^{-1}$ and dispersion of $\sigma=9.3$~km~s$^{-1}$. 

We then measured the velocities of each of the standards by two different methods of cross-correlation with the templates. In the first method, we split the spectrum into thirteen different 200~\AA~regions, spanning from 6400 to 9200~\AA. We cross-correlated each region with the template, requiring velocities less than $|$RV$|< 500$~km~s$^{-1}$. We then calculated the mean and standard deviation of the velocities and for those with standard deviations greater than 30~km~s$^{-1}$ iteratively rejected the farthest value from the mean. Our final velocities are measured from at least four regions and have standard deviations (adopted as the uncertainty) less than 30~km~s$^{-1}$. The middle panel shows measured velocities for ten of the 12 standards, which have a mean RV difference of $\Delta$RV=$-4.4$~km~s$^{-1}$ and dispersion of $\sigma=7.5$~km~s$^{-1}$. This cross-correlation method measured more velocities at slightly greater accuracy than the atomic line method, indicating that templates have an RV accuracy and precision of $\sim$10~km~$s^{-1}$. 

The second method is meant to extend to L dwarfs with lower S/N (e.g., the remaining two standards) by cross-correlating with a single, larger region of the spectrum (7000--9000~\AA). The velocities of all twelve standards are shown in the left panel of Figure~\ref{fig:knownv}. Overall, they have a mean RV difference of $\Delta$RV=$-9.4$~km~s$^{-1}$ and dispersion of $\sigma=15.1$~km~s$^{-1}$. While the velocities are both less accurate and precise, this single region cross-correlation measures radial velocities for lower S/N spectra, still with reasonable accuracy. 

To further test the accuracy and precision of both the cross-correlation methods with these spectra, we used a Monte Carlo approach. For each template, we generated 50000 simulated spectra with the wavelength array of an SDSS spectrum, a velocity shift, and added noise. We used the wavelength array from actual SDSS data to ensure an exact match to SDSS resolution. The velocity shifts were generated in a normal distribution with a mean of 0 and standard deviation of 34.4 km s$^{-1}$ (based on the RV distribution from S10). Noise was added using an array generated based on a normal distribution with a mean of 0 and standard deviations ranging from 0 to 1. We then measured the S/N of each generated spectrum using the standard deviation and mean in the region from 8200-8900\AA. We note that this measured S/N underestimates the true S/N of the spectrum because the complex structure of molecular bands increases the standard deviation (and thus, the measured ``noise"). 

\begin{figure*} 
\begin{center}
\includegraphics[width=0.4\linewidth]{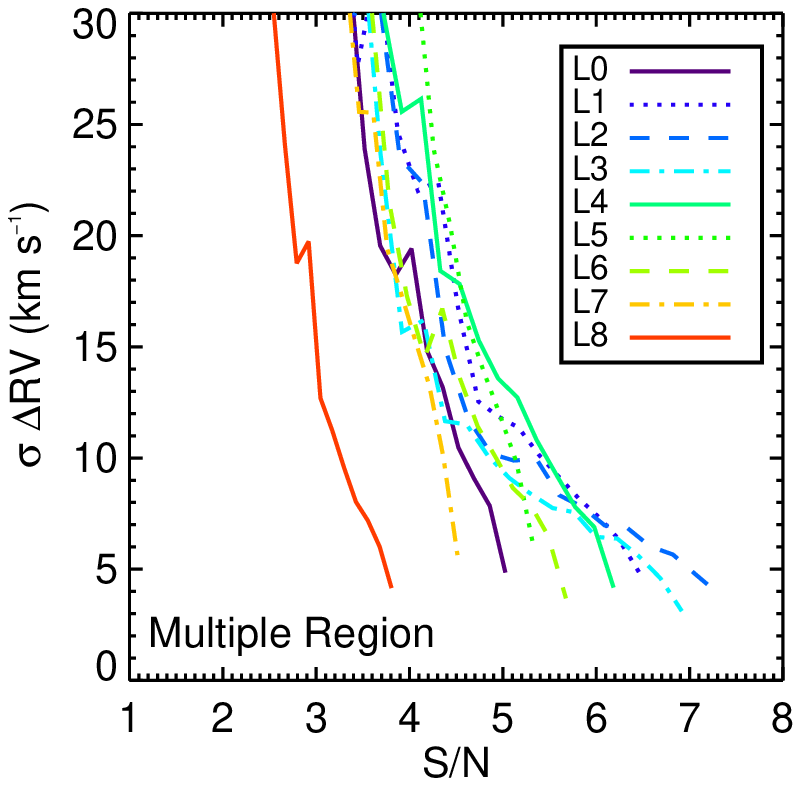} 
\includegraphics[width=0.4\linewidth]{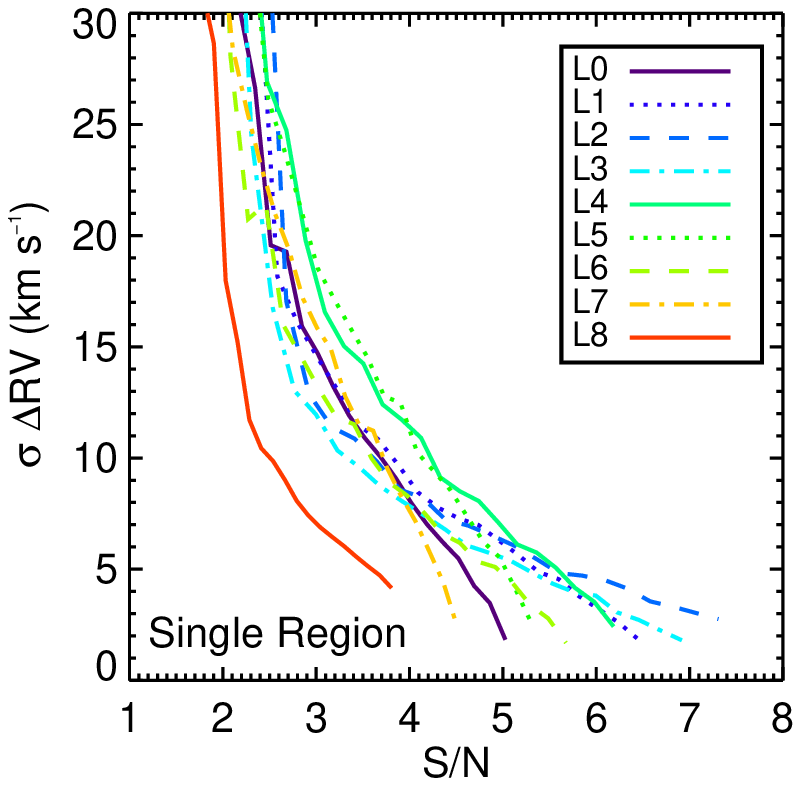}
\caption{The projected uncertainties of radial velocities measured against the template spectra as a function of S/N, as measured by the mean flux divided by the standard deviation in the 8200--8900\AA~region. The velocities in the left panel were measured using thirteen 200~\AA~wide regions from 6400 to 9200~\AA, and the velocities in the right panel were measured from cross-correlation over the 7000--9000\AA~range. The uncertainties are based on the standard deviation of the difference between simulated input and output radial velocities for each spectral type.}\label{fig:errc}
\end{center}
\end{figure*}

We then measured the RV of the simulated data using a cross-correlation between the spectrum and template both using the thirteen 200~\AA~regions (and applying the same iterative rejection) and using the entire 7000--9000\AA~region. The standard deviation of the difference between the measured RV and the input RV is shown as a function of S/N in Figure~\ref{fig:errc}. While the comparison to standards showed that using multiple regions results in increased precision and accuracy, these simulations indicate that a single, larger, cross-correlation region still results in reasonable accuracy for spectra with lower S/N. Both methods confirm the templates as suitable RV standards at the 10--20~km~s$^{-1}$ level. 

\subsection{$J-K_S$ Color and Optical Spectra}
\label{sec:jkl}
The spectra of ultracool dwarfs with redder than average $J-K_S$ colors often have features consistent with low gravity \citep[sharper atomic lines, weaker Na I, stronger VO, and weaker TiO;][]{Cruz2009}. These low-gravity, red L dwarfs are thought to have younger ages because they frequently have Li I detections and are often kinematically associated with young moving groups and open clusters \citep{Kirkpatrick2008}. Physically, these features correspond to lower surface gravity due to the larger radii of younger ultracool dwarfs \citep[e.g.,][]{Burrows1997}

In contrast, the optical spectra of ultracool dwarfs with blue $J-K_S$ colors are only mildly peculiar, and often in different ways. Some blue ultracool dwarfs are metal poor subdwarfs with strong TiO, CaH and VO bands \citep[due to less efficient dust cloud formation;][]{Burgasser2007,Burgasser2009}. Other blue ultracool dwarfs have thinner or patchier cloud structures than their red counterparts but less evidence of subdwarf features, such as SDSS 141624.08+134826.7 \citep{Bowler2010,Schmidt2010a}. In optical spectra, these thin clouds only cause slight spectral peculiarities (such as a narrower K I doublet). L dwarfs with red $J-K_S$ colors also have a smaller velocity dispersion than L dwarfs with blue $J-K_S$ colors, indicating that blue L dwarfs are likely to be an older population than the young, red L dwarfs \citep[][S10]{Faherty2009}.

\begin{figure*} 
\includegraphics[width=\linewidth]{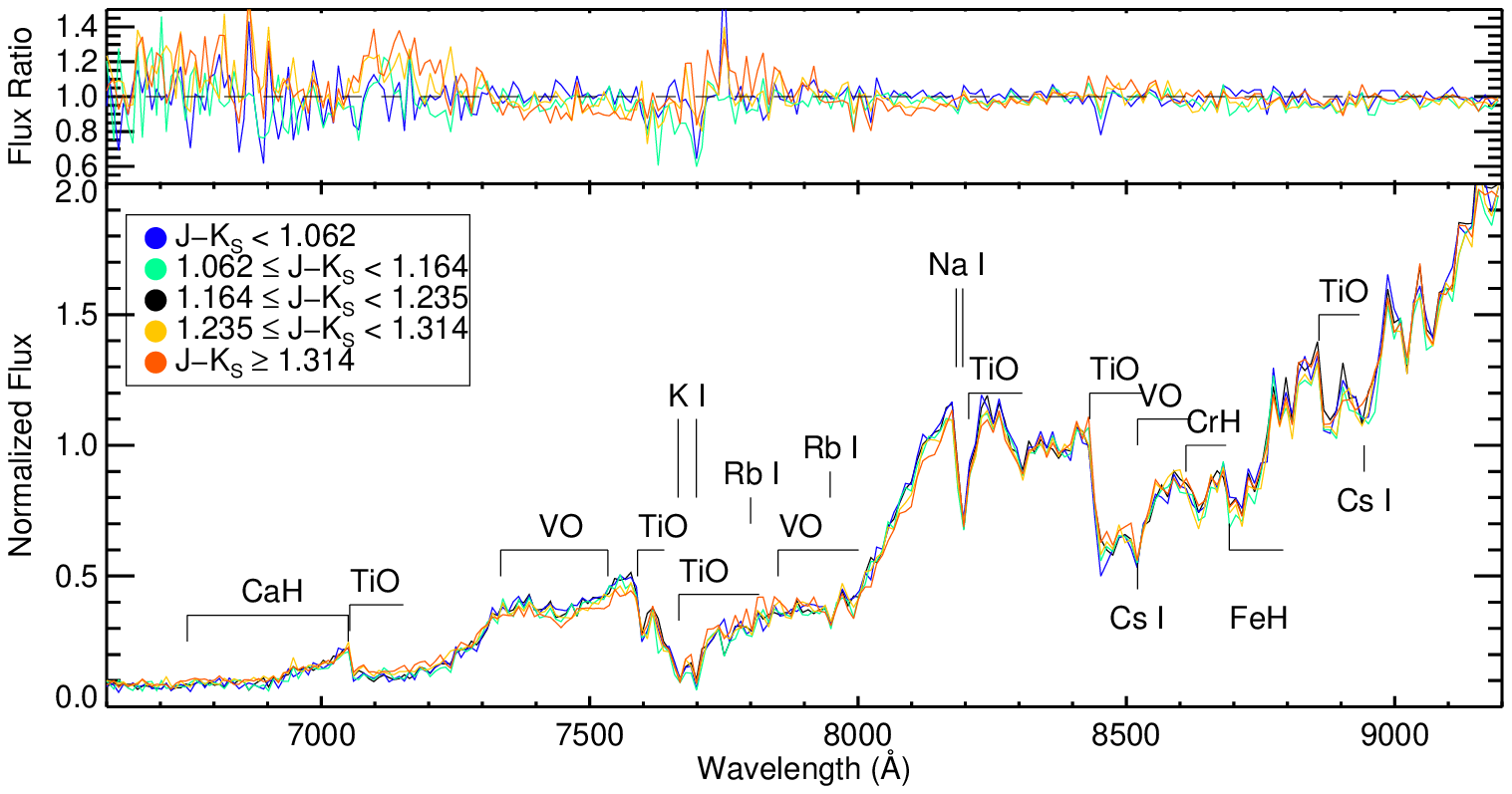} 
\caption{The bottom panel shows five L0 template spectra created from five bins based on their $J-K_S$ color (described in Section~\ref{sec:jkl}). In the top panel, red and blue template spectra are divided by the template spectrum of the color bin closest to the median $J-K_S$ color for L0 dwarfs (shown in black).}\label{fig:rvb}
\end{figure*}

To examine the optical spectra of L dwarfs with different $J-K_S$ colors, we divided the L0 subclass into five $J-K_S$ color bins with an equal ($\sim$60) number of objects. We constructed low-resolution templates from the spectra in each bin, using a velocity resolution of 400~km s$^{-1}$ ($\sim$10\AA) per pixel. Due to the low velocity resolution, we chose to include all L0 spectra (even those without reliable RVs) so we did not correct each spectrum to 0 km s$^{-1}$. These red and blue templates are shown in Figure~\ref{fig:rvb}. Some of the differences between the templates are small, therefore we also show the ratio of red and blue templates to template closest to the median $J-K_S$ color in the top panel of Figure~\ref{fig:rvb}.

While differences in atomic lines are not visible due to the low resolution of the templates, the reddest template shows weaker TiO and stronger VO than the median template, as does the next reddest template though with smaller differences. These features are especially strong for low-gravity L dwarfs \citep{Cruz2009}. Neither of the spectra in the blue bins differ significantly from the spectrum of the central $J-K_S$ bin. There is some evidence of a narrower K I doublet (the divided spectrum shows a stronger center and weaker wings). Even with the increased signal compared to single spectra, the blue templates show no increased ratio of CaH to TiO  \citep[common among mildly metal-poor M dwarfs, e.g.,][]{Gizis1997,Lepine2007} or features characteristic of the handful of known subdwarfs \citep[e.g.,][]{Burgasser2007}.

The features found in the red template spectra indicate that peculiarly red $J-K_S$ colors and low-gravity features are synonymous; most ultracool dwarfs with low-gravity spectra are red \citep{Cruz2009} and most peculiarly red ultracool dwarfs have some low-gravity features. The case is not the same for ultracool dwarfs with blue $J-K_S$ colors. Either there is no single cause (and thus no single effect on the optical spectra) of blue $J-K_S$ colors, or the root cause of blue $J-K_S$ colors does not affect the optical spectrum (either in the continuum or the lines) strongly enough to be observed in a median template spectrum.

\section{Bolometric Fluxes}
\label{sec:fbol}
We determined bolometric fluxes using a combination of the BT-Settl model spectra \citep{Allard2011}\footnote{Using \citet{Caffau2011} abundances; models from \url{http://phoenix.ens-lyon.fr/Grids/BT-Settl/CIFIST2011/}}, SDSS spectra spectrophotometrically calibrated to the $i$-band flux,  and the SDSS, 2MASS, and WISE photometry discussed in Section~\ref{sec:photdat}. Our basic procedure was to determine the best fit model spectrum for each object based on both the photometric and the spectroscopic data, then integrate over a combination of the model and observed spectra (normalized to the photometry) to calculate the bolometric flux. We further refined our calculation by interpolating between bolometric flux grid points spanning the best fit models. 

We required that each ultracool dwarf included in the bolometric flux calculations have good photometry (as defined by the uncertainty cuts noted in Section~\ref{sec:photdat}) in the $r$ through $W3$ bands for M7-M9 dwarfs, $r$ through $W2$ bands for L0 and L1 dwarfs, $i$ through $W2$ bands for L2 and L3 dwarfs, and $z$ through $W2$ bands for L4--L8 dwarfs. We used the $i$-band photometry for L4-L8 dwarfs even if the uncertainty was high because $i$-band photometry is required for spectrophotometric calibration of the SDSS spectrum. These cuts resulted in a sample of 108 ultracool dwarfs. 

We used a grid of 76 model spectra with a range of physical properties: $T_{\rm eff} = 1300$--$3100$~K, log(g) = 4.0 to 5.5, and constant [M/H]=0. For each model, we calculated model fluxes in each photometric band by integrating the model spectrum over the $r$, $i$, and $z$ filter curves from \citet{Doi2010}, the $J$, $H$, and $K_S$ filter curves from \citet{Cohen2003} and the $W1$, $W2$, and $W3$ filter curves from \citet{Wright2010}. Observed fluxes in each photometric band were calculated from the photometry of each ultracool dwarf using the zero-points and conversions in \citet{Lupton1999} for SDSS data, \citet{Cohen2003} for 2MASS data, and \citet{Jarrett2011} for WISE data. 

After the model and observed fluxes were calculated, we then determined a normalization constant\footnote{The dimensionless constant, $a$, represents the ratio of the radius of the ultracool dwarf to its distance. Due to the uncertainties in the fit to the model and the lack of parallaxes for these objects, we treat $a$ as a normalization constant.}, $a$, between the model and observed fluxes using the goodness-of-fit statistic:
\begin{equation}
 \frac{1}{n} \displaystyle\sum\limits_{i=0}^n \frac{(a \times F_{mod,i}-F_{data,i})^2}{\sigma_{data,i}^2},
\end{equation}
where $n$ is the number of bands with good photometry, $a$ is the model normalization, and $F_{mod,i}$, $F_{data,i}$, and $\sigma_{data,i}$ are the model flux, observed flux, and observed flux uncertainty in each photometric band ($i$). For each ultracool dwarf, we determined a best fit normalization, $a_{m}$, for each model based on the minimum of the goodness of fit statistic. 

We then calculated a grid of bolometric fluxes for each ultracool dwarf based on every unique combination of models and data. We first normalized the model spectrum to the observed photometric fluxes using $a_{mod}$. The model spectrum was then replaced by the spectrophotometrically calibrated SDSS spectrum between 6000 and 9200\AA~(for DR7 spectra) or 6000 and 10400\AA~(for DR10 spectra). We integrated over the combined model spectrum and SDSS spectrum between 10 and 10$^7$\AA~(the full range of the available models) to calculate a bolometric flux. The end result was a grid of bolometric fluxes for each ultracool dwarf that we examined as a function of model $T_{\rm eff}$ and log(g). Instead of assigning a bolometric flux based on a single, discrete best fit model, we performed a second fit to interpolate between points on the $T_{\rm eff}$ and log(g) grid.

For this last fit between the models and data, we calculated a goodness-of-fit statistic modified to use both the observed photometric fluxes and observed spectrum compared to the model fluxes and model spectrum:
\begin{equation}
\mbox{\tiny\( \frac{1}{n+1} \frac{1}{b_{tot}} \left( \displaystyle\sum\limits_{i=0}^n b_i \times \frac{(a_{m} \times F_{mod,i}-F_{data,i})^2}{\sigma_{data,i}^2} + \frac{b_{spec}}{m} \times \displaystyle\sum\limits_{j=0}^m \frac{(a_{m} \times F_{mod,j}-F_{spec,j})^2}{\sigma_{spec,j}^2} \right),\)}
\end{equation}
where, as above, there are $n$ bands of good photometry, but the new $j$ index iterates through $m$ spectroscopic data points, $F_{mod,j}$ is the model flux interpolated onto the SDSS spectrum wavelength array, $F_{spec,j}$ is the spectroscopic flux, and $\sigma_{spec,j}$ is the spectroscopic error. Each photometric flux point and the spectrum are normalized by $b_i$, the width of each photometric band. 

We calculated the best fit $T_{\rm eff}$ by finding the minimum of a polynomial fit to the goodness-of-fit as a function of $T_{\rm eff}$ with the best fit log(g) held constant. The uncertainties on the $T_{\rm eff}$ were assigned where the goodness-of-fit statistic was equal to its minimum plus one \citep{Bevington1969}. Since there were too few log(g) values to fit a polynomial, we adopted the best fit value and used the entire range of log(g) (4.0 to 5.5) as the uncertainty. 

We then interpolated the bolometric flux grid between the points to log(g) and $T_{\rm eff}$ to assign the final bolometric flux and uncertainty. We rejected seven dwarfs with total uncertainties larger than 50\% of their bolometric flux, resulting in bolometric fluxes for 101 dwarfs, which are given in Table~\ref{tab:fbol}. 

\setcounter{table}{4}

\begin{figure*}
\includegraphics[width=0.95\linewidth]{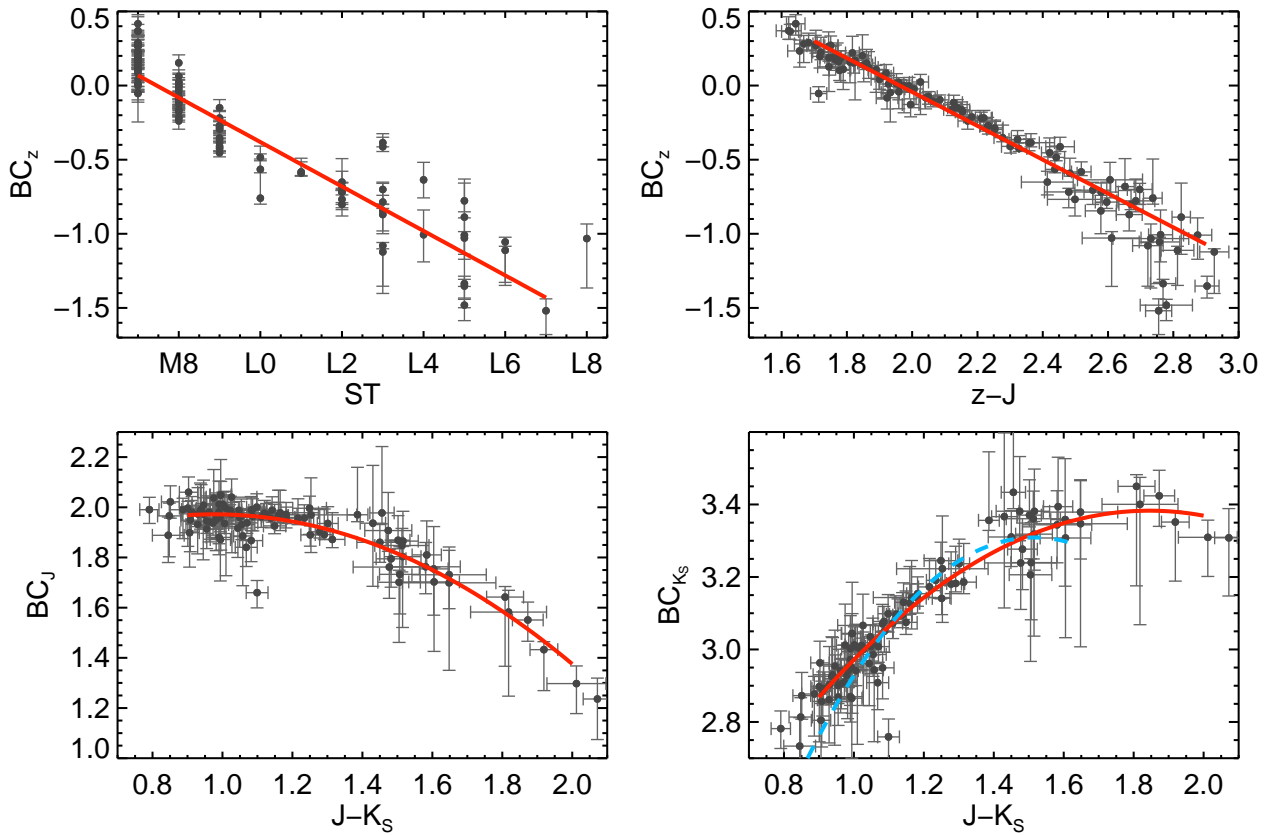} 
\caption{Bolometric corrections in the $z$-band as a function of spectral type (top left panel) and $z-J$ color (top right panel), bolometric corrections in the $J$-band as a function of $J-K_S$ color (bottom left panel), and bolometric corrections in the $K_S$ band as a function of $J-K_S$ color (bottom right panel). The coefficients for the polynomial and linear relations (red lines) are given in Table~\ref{tab:BC}. In the bottom right panel, the BC$_K$ relation from \citet{Leggett2001} has been converted to the 2MASS magnitude system and is shown (blue dashed line) for comparison.} \label{fig:BC}
\end{figure*}

We converted the bolometric fluxes to apparent bolometric magnitudes and calculated bolometric corrections in the $z$, $J$, and $K_S$ bands. The mean and median of each correction as a function of spectral type are given in Table~\ref{tab:chibolst}. While the bolometric corrections are correlated with spectral type, there is a significant amount of scatter compared to the dynamic range of each bolometric correction. The tightest relations between color and bolometric corrections are in $z$-band as a function of $z-J$ color and $J$- and $K_S$- bands as a function of $J-K_S$ color. We show BC$_z$ as a function of spectral type in addition to those three relations in Figure~\ref{fig:BC} and give polynomial fits in Table~\ref{tab:BC}. For comparison, we converted the \citet{Leggett2001} relationship between BC$_{K}$ and $J-K$ color in the MKO system to 2MASS colors \citep[based on the relations from][]{Carpenter2001}. The \citet{Leggett2001} BC$_{K}$ shows good agreement with the data near $J-K_S=1.4$, but under-predicts BC$_{Ks}$ and near $J-K_S=1.6$ (likely due to the absence of objects redder than $J-K_S>1.6$ in that dataset). On the blue end, the data from \citet{Leggett2001} show significant scatter, which is broadly consistent with both fits. The data from this paper extend to redder $J-K_S$ colors, allowing a fit that extends past the \citet{Leggett2001} fit. 

\begin{deluxetable*}{l|l|lll|lll|lll|llll} \tablewidth{0pt}
\tablecaption{Bolometric Corrections and $\chi$ values For Each Spectral Type\label{tab:chibolst}}
\tablehead{
\colhead{Spectral} & \colhead{BC} & \multicolumn{3}{c}{BC$_z$} & \multicolumn{3}{c}{BC$_J$} & \multicolumn{3}{c}{BC$_{Ks}$} &  \multicolumn{4}{c}{$\chi$ value ($\times 10^{-6}$)} \\ \colhead{Type} & \colhead{\#} & \colhead{median} & \colhead{mean} & \colhead{$\sigma$} & \colhead{median} & \colhead{mean} & \colhead{$\sigma$} & \colhead{median} & \colhead{mean} & \colhead{$\sigma$} & \colhead{\#} &  \colhead{median} & \colhead{mean} & \colhead{$\sigma$} }
\startdata
M7 & 36 & \phs0.18 & \phs0.17 & 0.11 & 1.97 & 1.96 & 0.07 & 2.92 & 2.91 & 0.08 & 36 & 10.28 & 10.64 & 3.13 \\ 
M8 & 21 & $-$0.08 & $-$0.08 & 0.10 & 1.97 & 1.96 & 0.05 & 3.00 & 3.00 & 0.07 & 21 & 4.26 & 4.43 & 1.18 \\ 
M9 & 12 & $-$0.30 & $-$0.32 & 0.09 & 1.96 & 1.95 & 0.03 & 3.13 & 3.12 & 0.04 & 11 & 2.52 & 2.69 & 0.58 \\ 
L0 & 3 & $-$0.57 & $-$0.60 & 0.14 & 1.96 & 1.94 & 0.06 & 3.19 & 3.26 & 0.15 & 3 & 1.98 & 2.00 & 0.27 \\ 
L1 & 2 & $-$0.58 & $-$0.59 & 0.01 & 1.94 & 1.91 & 0.03 & 3.24 & 3.21 & 0.04 & 2 & 2.25 & 2.18 & 0.11 \\ 
L2 & 7 & $-$0.72 & $-$0.72 & 0.05 & 1.85 & 1.83 & 0.08 & 3.36 & 3.34 & 0.05 & 3 & 2.11 & 2.18 & 0.36 \\ 
L3 & 8 & $-$0.79 & $-$0.78 & 0.27 & 1.81 & 1.85 & 0.14 & 3.28 & 3.27 & 0.12 & 4 & 1.67 & 1.64 & 0.22 \\ 
L4 & 2 & $-$0.64 & $-$0.82 & 0.26 & 1.97 & 1.86 & 0.15 & 3.36 & 3.29 & 0.10 & 1 & 1.16 & 1.16 & \nodata \\ 
L5 & 7 & $-$1.03 & $-$1.12 & 0.27 & 1.58 & 1.65 & 0.25 & 3.38 & 3.37 & 0.04 & 4 & 1.46 & 1.25 & 0.28 \\ 
L6 & 2 & $-$1.06 & $-$1.08 & 0.04 & 1.70 & 1.70 & 0.00 & 3.31 & 3.26 & 0.07 & 1 & 1.23 & 1.23 & \nodata \\ 
L7 & 1 & $-$1.52 & $-$1.52 & \nodata & 1.24 & 1.24 & \nodata & 3.31 & 3.31 & \nodata & 1 & 0.73 & 0.73 & \nodata \\ 
L8 & 1 & $-$1.03 & $-$1.03 & \nodata & 1.70 & 1.70 & \nodata & 3.35 & 3.35 & \nodata & 0 & \nodata & \nodata & \nodata
\enddata
\end{deluxetable*}

\begin{deluxetable*}{lllllll} \tablewidth{0pt} \tabletypesize{\scriptsize} 
\tablecaption{Bolometric Correction Polynomial Relations \label{tab:BC}}
\tablehead{\colhead{ } & \colhead{color}  & \colhead{$a_1$}   & \colhead{$a_2$} & \colhead{$a_3$}    &  \colhead{Range}  & \colhead{$\sigma$\tablenotemark{a}}  }
\startdata
BC$_z$ & ST\tablenotemark{b}  & 1.12    &  $-0.150 $ & \nodata  &     $ 7 \leq {\rm ST} \leq 17$ & 0.156 \\
BC$_z$ & $z-J$            &    2.24   &  $-1.14$  &  \nodata  &  $1.7 < z-J < 2.9$ & 0.148 \\
BC$_J$ & $J-K_S$      &   1.43   &  \phs1.11 & $-0.567$  & $0.9 < J-K_S < 2$ & 0.569 \\
BC$_{Ks}$ & $J-K_S$ &   1.42  &  \phs2.13 & $-0.576$ & $0.9 < J-K_S < 2$ & 0.536
\enddata
\tablecomments{For a polynomial of the form ${\rm BC}=a_1 + a_2x + a_3x^2$ where $x$ is the color/spectral type.}
\tablenotetext{a}{$\sigma$ is the mean error on $y$ based on the coefficient uncertainties.}
\tablenotetext{b}{Where M7 = 7, L0 = 10, and L8 = 18.}
\end{deluxetable*}

\section{$\chi$ Values}
\label{sec:chi}
\begin{figure}
\includegraphics[width=\linewidth]{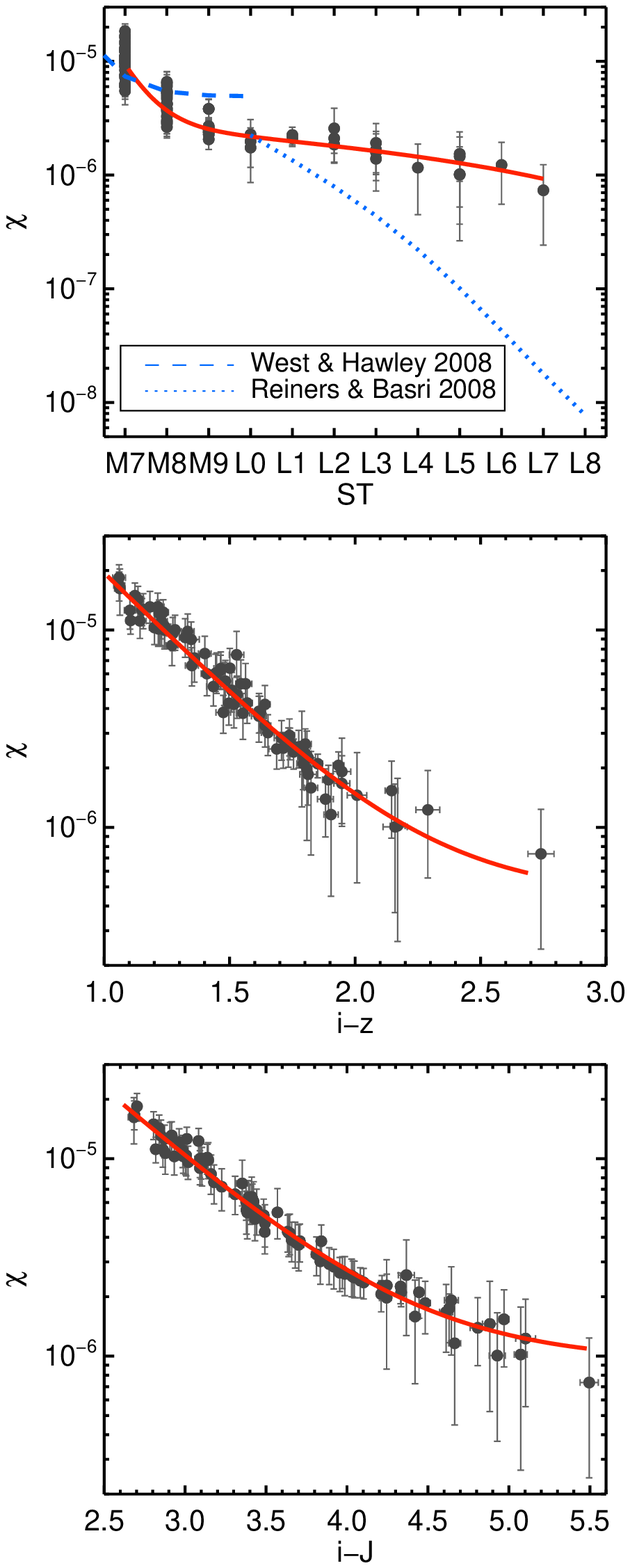} 
\caption{The $\chi$ values for ultracool dwarfs (see Section~\ref{sec:chi}) as a function of spectral type (top panel), $i-z$ color (middle panel) and $i-J$ color (bottom panel). The top panel shows comparison with previous results from \citet{West2008a} and \citet{Reiners2008}, shown with blue dashed and dotted lines respectively. Fits are shown in each panel (red solid lines).} \label{fig:chi}
\end{figure}

As discussed in Section~\ref{sec:intro}, the $\chi$ value is the ratio of the flux in the H$\alpha$ continuum to the bolometric flux, used to convert H$\alpha$ EW to $L_{\rm H\alpha}/L_{\rm bol}$. The H$\alpha$ continuum was measured from SDSS spectra; we first spectrophotometrically calibrated each spectrum to its SDSS $i$-band magnitude, then measured the mean continuum value using the H$\alpha$ continuum regions from the Hammer spectral typing software \citep[6500-6550\AA~and 6575-6625\AA;][]{West2004,Covey2007}. The $\chi$ value is then the ratio of the mean continuum value to the bolometric fluxes described in Section~\ref{sec:fbol}. We adopted the standard deviation of the continuum region as the uncertainty in the continuum flux and calculated a final uncertainty on $\chi$ ($\sigma_{\chi}$) by propagating the mean bolometric flux uncertainties and the continuum uncertainties. We excluded dwarfs with $\chi$ uncertainties of $\sigma_{\chi} > 0.75 \chi$, resulting in a final sample of 87 objects. The calculated $\chi$ values for each ultracool dwarf are given in Table~\ref{tab:fbol} and shown as a function of spectral type, $i-z$ color, and $i-J$ color in Figure~\ref{fig:chi}. 

We provide fits for $\chi$ as a function of spectral type, $i-z$, and $i-J$ (Table~\ref{tab:echi}). To fit $\chi$ as a function of spectral type and color, \citet{West2008a} use a modified exponential equation of the form $\chi = ae^{-x/b}+c$. We adopt that equation, but add a linear component to the spectral type relation that is needed to fit the mid- to late-L $\chi$ as a function of spectral type: $\chi = ae^{-x/b} + cx + d$. The spread in values for spectral type are much higher than those for the colors, but we provide a spectral type relation because SDSS does not have full sky coverage and there is a significant spread in any relation between $\chi$ and 2MASS/WISE colors. 

\begin{deluxetable*}{lllllll} \tablewidth{0pt} \tabletypesize{\scriptsize} 
\tablecaption{Fits to $\chi$ \label{tab:echi}}
\tablehead{\colhead{x} & \colhead{$a$}  & \colhead{$b$}   & \colhead{$c$ ($\times10^{-7}$)}  & \colhead{$d$ ($\times10^{-6}$)} & \colhead{Range}  & \colhead{$\sigma$\tablenotemark{a} ($\times10^{-6}$)}  }
\startdata
$i-z$ &  $3.50  \times10^{-4}$ & $0.344 $ & \nodata & $0.448 $   & 1 $< i-z  <$ 2.5 & $0.642$  \\
$i-J$ &   $1.39 \times 10^{-3}$ & $0.602 $ & \nodata & $0.939 $   & 2.6 $< i-J < $ 5.5 & $0.499$ \\
ST \tablenotemark{b}    &  $1.09  $ & $0.584 $ & $-1.74 $ & $3.88 $  & 7 $<$ ST $<$ 17 & $1.43$
\enddata
\tablecomments{Where $\chi = ae^{-x/b}+ cx + d$ where x is the color/spectral type.}
\tablenotetext{a}{$\sigma$ is the mean error on $y$ based on the coefficient uncertainties.}
\tablenotetext{b}{For M7 = 7, L0 = 10, and L8 = 18}
\end{deluxetable*}

\begin{figure}
\begin{center}
\includegraphics[width=0.9\linewidth]{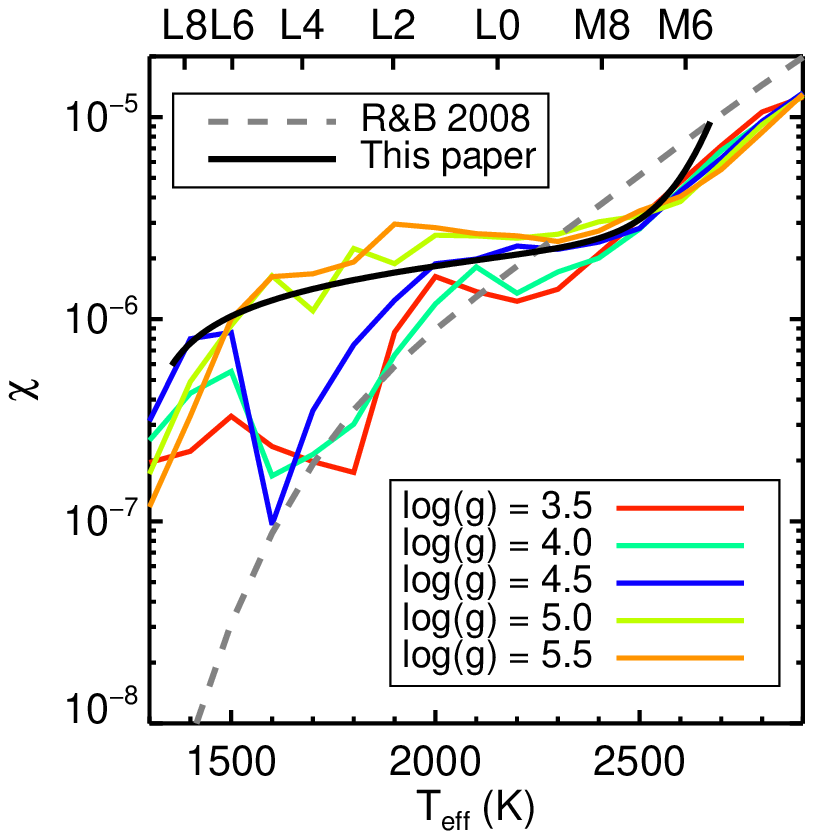} 
\includegraphics[width=0.9\linewidth]{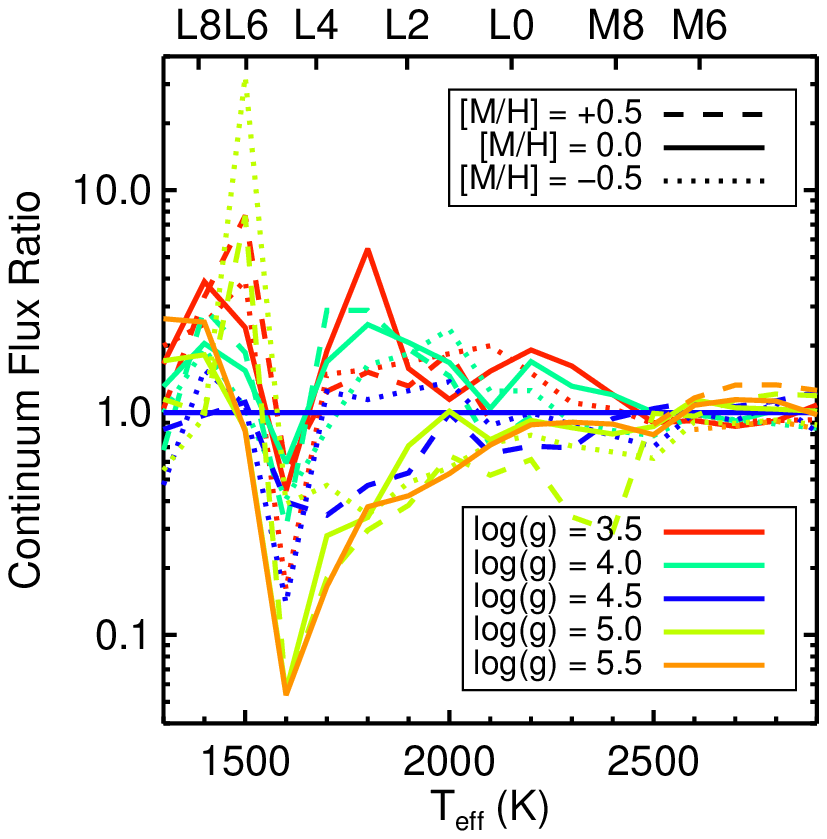} 
\caption{In the left panel, the $\chi$ values calculated from the earlier version of the PHOENIX models from \citet[][grey dashed line]{Reiners2008} shown with the $\chi$ values calculated from the BT-Settl grid of models (colored lines). The $\chi$ as a function of spectral type derived in this paper is also shown (solid black line); the conversion between spectral type and $T_{\rm eff}$ is from \citet{Stephens2009}. The different log(g) values for the BT-Settl results are given in the lower right corner. In the right panel, the ratio of the continuum flux of the BT-Settl model spectra for a range of [M/H] (top right corner) and log(g) (bottom right corner) compared to the [M/H] = 0.0 and log(g) = 4.5 model for the same $T_{\rm eff}$.} \label{fig:modchi}
\end{center}
\end{figure}

The newly calculated $\chi$ value shows reasonable agreement with the \citet{West2008a} $\chi$ relation. The two completely agree at M7, then diverge for M8 and M9, likely due to the small number of late-M dwarfs available in their study. There is a much larger difference between our $\chi$ values and those calculated from \citet{Reiners2008}. Their values are based on the DUSTY set of models \citep{Allard2001}, while our new $\chi$ values are both based on a different model grid (BT-Settl) and use spectroscopic data to measure the continuum around H$\alpha$. To better understand the differences between the two model grids and the calculated $\chi$ values, we examine the \citet{Reiners2008} $\chi$ values compared to a $\chi$ value calculated directly from the BT-Settl model spectra and our new $\chi$ value in Figure~\ref{fig:modchi}.  We converted our $\chi$ as a function of spectral type to $T_{\rm eff}$ using the \citet{Stephens2009} conversion.

The difference between our new $\chi$ values and the \citet{Reiners2008} $\chi$ relation is minor for $T_{\rm eff} = 2000$--$3000$~K (spectral types M7-L1), but increases to almost an order of magnitude at $T_{\rm eff}=1500$~K. The BT-Settl models, shown for a range of log(g), span the difference between the two relations, with log(g)$=3.5$--$4.5$ showing agreement with the \citet{Reiners2008} $\chi$ relation down to $T_{\rm eff}=1600$~K and the log(g)$=5.0$--$5.5$ showing agreement with our semi-empirical $\chi$ value over the full range of $T_{\rm eff}$. The agreement of the lower gravity BT-Settl model $\chi$ with the DUSTY model $\chi$ is probably due to the similarities between the two sets of models; the DUSTY models have a relatively thick clouds and are most similar to the spectra of low gravity objects. 

As discussed in Section~\ref{sec:describe_mod}, $F_{bol}$ varies by $\sim$10\% with [M/H] or log(g), so the order of magnitude differences are primarily due to variations in continuum surrounding H$\alpha$. To investigate variations in the continuum, we measured the continuum surrounding H$\alpha$ in the BT-Settl grid spanning $T_{\rm eff} = 1300$--$3100$~K (in steps of 100~K), log(g)$= 3.5$--$5.5$ (in steps of 0.5 dex), and $-0.5$, $0.0$, and $+0.5$. The ratio of the continuum flux for each model spectrum compared to he [M/H] = 0.0 and log(g) = 4.5 model for the same $T_{\rm eff}$ is shown in the right panel of Figure~\ref{fig:modchi}.

The H$\alpha$ continuum strongly depends on both metallicity and gravity at $T_{\rm eff}<2500$~K, with variations spanning three orders of magnitude. While individual models may deviate, higher gravity and metallicity generally result in suppressed continuum flux. At these temperatures, the main contributors to opacity around the H$\alpha$ line are the pressure broadened $\lambda$5890,5896\AA~Na I and $\lambda$7665,7699\AA~K I doublets. Higher gravity increases the pressure broadening and thus the opacity far from the line center; higher metallicity increases the abundance of the absorbing species. While the specific log(g) and [M/H] distributions of the sample of late-M and L dwarfs used to measure the $\chi$ value are unknown, the sample is drawn mainly from the field population (as evidenced by its kinematics; e.g., S10), so the $\chi$ values determined are likely to be most accurate for field ultracool dwarfs. Especially young ultracool dwarfs \citep[cloudy, low gravity objects, e.g.,][]{Cruz2009} or those with lower metallicities \citep[ultracool subdwarfs, e.g.,][]{Burgasser2007} may have systematically lower $\chi$ values.

\section{Summary}
We have presented new templates for L dwarfs and updated bolometric corrections and $\chi$ values for late-M and L dwarfs. The L dwarf spectroscopic templates are suitable for use both as spectral standards and as RV standards with precision ranging from 10 to 20 km s$^{-1}$. Using templates produced from spectra of dwarfs that were separated into red and blue bins based on their $J-K_S$ colors, we have shown that most red L dwarfs show spectral features consistent with lower surface gravity in their optical spectra, while blue L dwarfs show no spectral features that deviate strongly from the template of L dwarfs in the median color bin. Optical spectra are poor discriminants of the physical properties that are suspected to correspond with blue $J-K_S$ color: mildly sub-solar metallicities or patchy clouds.

We determined new bolometric fluxes and bolometric corrections based on models, photometry, and spectra of 101 ultracool dwarfs. The bolometric corrections (accurate to 10--20\%) are consistent with previous values. The new $\chi$ values are nearly two orders of magnitude larger than previous $\chi$ values for late-L dwarfs. The models used for previous $\chi$ measurements were better matches for younger, lower surface gravity ultracool dwarfs and systematically underestimated the continuum for stars closer to typical field ages. Our $\chi$ values are likely to be most accurate for field ultracool dwarfs, and may overestimate $\chi$ for low gravity and low metallicity objects by up to an order of magnitude. Updated activity strength for L dwarfs in the field will be discussed in Schmidt et al. (2014, in prep.).

\setcounter{table}{3}

\acknowledgments
A.\ A.\ W.\ acknowledges funding from NSF grants AST-1109273 and AST-1255568 and also the support of the Research Corporation for Science Advancement's Cottrell Scholarship.

This research has benefitted from the M, L, and T dwarf compendium housed at \url{DwarfArchives.org} and maintained by Chris Gelino, Davy Kirkpatrick, and Adam Burgasser. This research has also benefitted from the SpeX Prism Spectral Libraries, maintained by Adam Burgasser at \url{http://pono.ucsd.edu/$\sim$adam/browndwarfs/spexprism}.

This publication makes use of data products from the Two Micron All Sky Survey, which is a joint project of the University of Massachusetts and the Infrared Processing and Analysis Center/California Institute of Technology, funded by the National Aeronautics and Space Administration and the National Science Foundation.

This publication also makes use of data products from the Wide-field Infrared Survey Explorer, which is a joint project of the University of California, Los Angeles, and the Jet Propulsion Laboratory/California Institute of Technology, funded by the National Aeronautics and Space Administration.

Funding for the SDSS and SDSS-II has been provided by the Alfred P. Sloan Foundation, the Participating Institutions, the National Science Foundation, the U.S. Department of Energy, the National Aeronautics and Space Administration, the Japanese Monbukagakusho, the Max Planck Society, and the Higher Education Funding Council for England. The SDSS Web Site is \url{http://www.sdss.org/}.

The SDSS is managed by the Astrophysical Research Consortium for the Participating Institutions. The Participating Institutions are the American Museum of Natural History, Astrophysical Institute Potsdam, University of Basel, University of Cambridge, Case Western Reserve University, University of Chicago, Drexel University, Fermilab, the Institute for Advanced Study, the Japan Participation Group, Johns Hopkins University, the Joint Institute for Nuclear Astrophysics, the Kavli Institute for Particle Astrophysics and Cosmology, the Korean Scientist Group, the Chinese Academy of Sciences (LAMOST), Los Alamos National Laboratory, the Max-Planck-Institute for Astronomy (MPIA), the Max-Planck-Institute for Astrophysics (MPA), New Mexico State University, Ohio State University, University of Pittsburgh, University of Portsmouth, Princeton University, the United States Naval Observatory, and the University of Washington.

\clearpage

\tabletypesize{\tiny}
\LongTables

\begin{deluxetable*}{llllllclcl} \tablewidth{0pt} \tabletypesize{\scriptsize} 
\tablecaption{Bolometric Fluxes For Ultracool Dwarfs \label{tab:fbol}}
\tablehead{\colhead{Designation} & \colhead{ST} & \colhead{$i$} & \colhead{$i-z$} & \colhead{$i-J$} & \colhead{$J-K_S$} & \colhead{$F_{bol}$} & \colhead{$m_{bol}$\tablenotemark{a}} & \colhead{$F_{H\alpha,cont}$} & \colhead{$\log(\chi)$} \\ \colhead{} &  \colhead{} &  \colhead{} &  \colhead{} &  \colhead{} &  \colhead{} & \colhead{[10$^{-11}$ erg s$^{-1}$ cm$^{-2}$]} &  \colhead{} &  \colhead{[10$^{-16}$ erg s$^{-1}$ cm$^{-2}$ \AA$^{-1}$]} &  \colhead{} }
\startdata
001309.3$-$002552.0 & M7 & 15.27 $\pm$ 0.02 & 1.32 $\pm$ 0.02 & 3.10 $\pm$ 0.03 & 0.85 $\pm$ 0.03 & $5.68^{+0.25}_{-0.78}$ & 14.1 & 5.20 $\pm$ 1.1 & $-$5.04 \\
012431.2$-$002756.3 & M7 & 15.25 $\pm$ 0.02 & 1.33 $\pm$ 0.03 & 3.14 $\pm$ 0.03 & 0.98 $\pm$ 0.03 & $5.99^{+0.22}_{-0.72}$ & 14.1 & 5.89 $\pm$ 1.1 & $-$5.01 \\
031225.1+002158.3 & M7 & 15.23 $\pm$ 0.02 & 1.23 $\pm$ 0.03 & 3.00 $\pm$ 0.03 & 1.01 $\pm$ 0.03 & $5.52^{+0.36}_{-0.17}$ & 14.2 & 5.61 $\pm$ 0.95 & $-$4.99 \\
033035.1$-$002535.7 & L5 & 20.09 $\pm$ 0.03 & 2.09 $\pm$ 0.04 & 4.77 $\pm$ 0.06 & 1.47 $\pm$ 0.07 & $0.331^{+0.047}_{-0.020}$ & 17.2 & \nodata & \nodata \\
035100.0$-$005245.8 & M7 & 14.69 $\pm$ 0.02 & 1.47 $\pm$ 0.03 & 3.39 $\pm$ 0.03 & 1.07 $\pm$ 0.03 & $12.9^{+0.65}_{-0.43}$ & 13.2 & 8.25 $\pm$ 1.7 & $-$5.19 \\
053503.5$-$001511.7 & M7 & 16.31 $\pm$ 0.01 & 1.11 $\pm$ 0.02 & 2.82 $\pm$ 0.02 & 1.10 $\pm$ 0.03 & $2.22^{+0.091}_{-0.11}$ & 15.2 & 2.48 $\pm$ 0.35 & $-$4.95 \\
054004.4+002128.6 & M7 & 17.42 $\pm$ 0.01 & 1.06 $\pm$ 0.02 & 2.68 $\pm$ 0.04 & 0.94 $\pm$ 0.06 & $0.523^{+0.047}_{-0.026}$ & 16.7 & 0.848 $\pm$ 0.21 & $-$4.79 \\
074642.4+200031.8 & L1 & 16.09 $\pm$ 0.01 & 1.85 $\pm$ 0.02 & 4.33 $\pm$ 0.02 & 1.29 $\pm$ 0.03 & $8.85^{+0.86}_{-0.095}$ & 13.7 & 1.86 $\pm$ 0.22 & $-$5.68 \\
074653.2+394509.9 & M7 & 15.51 $\pm$ 0.01 & 1.22 $\pm$ 0.02 & 2.93 $\pm$ 0.02 & 0.95 $\pm$ 0.03 & $4.10^{+0.30}_{-0.28}$ & 14.5 & 4.21 $\pm$ 0.76 & $-$4.99 \\
080255.7+332143.6 & L2 & 20.06 $\pm$ 0.03 & 1.71 $\pm$ 0.04 & 4.21 $\pm$ 0.09 & 1.65 $\pm$ 0.11 & $0.238^{+0.012}_{-0.023}$ & 17.6 & \nodata & \nodata \\
081058.6+142038.1 & M8 & 16.62 $\pm$ 0.01 & 1.74 $\pm$ 0.02 & 3.89 $\pm$ 0.02 & 1.14 $\pm$ 0.03 & $3.30^{+0.16}_{-0.13}$ & 14.7 & 0.965 $\pm$ 0.20 & $-$5.53 \\
082111.9+094946.0 & M7 & 15.93 $\pm$ 0.02 & 1.21 $\pm$ 0.02 & 2.92 $\pm$ 0.03 & 0.91 $\pm$ 0.03 & $2.58^{+0.15}_{-0.088}$ & 15.0 & 3.38 $\pm$ 0.57 & $-$4.88 \\
082519.4+211550.2 & L7 & 20.60 $\pm$ 0.05 & 2.74 $\pm$ 0.05 & 5.50 $\pm$ 0.06 & 2.07 $\pm$ 0.04 & $0.746^{+0.055}_{-0.10}$ & 16.3 & 0.0548 $\pm$ 0.037 & $-$6.13 \\
082906.6+145620.7 & L3 & 19.17 $\pm$ 0.02 & 1.82 $\pm$ 0.03 & 4.42 $\pm$ 0.03 & 1.58 $\pm$ 0.04 & $0.607^{+0.079}_{-0.021}$ & 16.6 & 0.0963 $\pm$ 0.051 & $-$5.80 \\
083557.3+431831.1 & M8 & 16.54 $\pm$ 0.02 & 1.50 $\pm$ 0.02 & 3.49 $\pm$ 0.03 & 1.08 $\pm$ 0.03 & $2.76^{+0.090}_{-0.20}$ & 14.9 & 1.18 $\pm$ 0.26 & $-$5.37 \\
084106.8+603506.3 & L4 & 20.40 $\pm$ 0.04 & 1.85 $\pm$ 0.05 & 4.45 $\pm$ 0.10 & 1.25 $\pm$ 0.13 & $0.175^{+0.019}_{-0.018}$ & 17.9 & \nodata & \nodata \\
084816.6+330908.6 & M7 & 15.17 $\pm$ 0.01 & 1.18 $\pm$ 0.03 & 2.85 $\pm$ 0.02 & 0.91 $\pm$ 0.03 & $4.99^{+0.25}_{-0.54}$ & 14.3 & 6.53 $\pm$ 1.1 & $-$4.88 \\
084900.5+022015.5 & M8 & 16.34 $\pm$ 0.01 & 1.50 $\pm$ 0.02 & 3.41 $\pm$ 0.02 & 1.03 $\pm$ 0.03 & $2.78^{+0.11}_{-0.37}$ & 14.9 & 1.78 $\pm$ 0.40 & $-$5.19 \\
085224.6+254058.8 & M7 & 15.33 $\pm$ 0.02 & 1.22 $\pm$ 0.02 & 3.14 $\pm$ 0.03 & 1.01 $\pm$ 0.03 & $5.35^{+0.29}_{-0.17}$ & 14.2 & 5.41 $\pm$ 0.90 & $-$4.99 \\
085938.5+634133.5 & L0 & 17.95 $\pm$ 0.01 & 1.81 $\pm$ 0.02 & 4.24 $\pm$ 0.03 & 1.31 $\pm$ 0.04 & $1.51^{+0.15}_{-0.027}$ & 15.6 & 0.298 $\pm$ 0.17 & $-$5.70 \\
090206.9+003319.4 & M8 & 15.53 $\pm$ 0.01 & 1.45 $\pm$ 0.02 & 3.42 $\pm$ 0.02 & 0.94 $\pm$ 0.03 & $5.96^{+0.19}_{-0.92}$ & 14.1 & 3.61 $\pm$ 0.76 & $-$5.22 \\
091130.5+224810.7 & M7 & 14.73 $\pm$ 0.01 & 1.23 $\pm$ 0.02 & 2.99 $\pm$ 0.02 & 0.98 $\pm$ 0.02 & $8.51^{+0.58}_{-0.53}$ & 13.7 & 9.43 $\pm$ 1.6 & $-$4.96 \\
091614.9+213950.8 & M9 & 17.43 $\pm$ 0.01 & 1.93 $\pm$ 0.02 & 4.21 $\pm$ 0.03 & 1.15 $\pm$ 0.03 & $2.23^{+0.18}_{-0.047}$ & 15.1 & 0.459 $\pm$ 0.078 & $-$5.69 \\
092125.3+601747.2 & M9 & 17.30 $\pm$ 0.01 & 1.73 $\pm$ 0.02 & 3.99 $\pm$ 0.02 & 1.16 $\pm$ 0.03 & $1.97^{+0.092}_{-0.041}$ & 15.3 & 0.515 $\pm$ 0.11 & $-$5.58 \\
092825.5+423053.6 & M9 & 17.05 $\pm$ 0.02 & 1.74 $\pm$ 0.02 & 3.97 $\pm$ 0.03 & 1.14 $\pm$ 0.03 & $2.47^{+0.14}_{-0.068}$ & 15.0 & 0.666 $\pm$ 0.12 & $-$5.57 \\
093113.2+280227.1 & L3 & 19.62 $\pm$ 0.03 & 1.95 $\pm$ 0.04 & 4.64 $\pm$ 0.04 & 1.25 $\pm$ 0.05 & $0.416^{+0.014}_{-0.036}$ & 17.0 & 0.0797 $\pm$ 0.037 & $-$5.72 \\
093126.0+363821.6 & M7 & 16.69 $\pm$ 0.01 & 1.48 $\pm$ 0.02 & 3.38 $\pm$ 0.03 & 0.96 $\pm$ 0.03 & $2.03^{+0.13}_{-0.12}$ & 15.2 & 1.12 $\pm$ 0.27 & $-$5.26 \\
093128.2+052821.9 & M8 & 16.30 $\pm$ 0.02 & 1.51 $\pm$ 0.03 & 3.43 $\pm$ 0.03 & 1.07 $\pm$ 0.03 & $3.35^{+0.055}_{-0.22}$ & 14.7 & 1.66 $\pm$ 0.37 & $-$5.31 \\
093220.8+391612.2 & M8 & 17.19 $\pm$ 0.02 & 1.64 $\pm$ 0.02 & 3.81 $\pm$ 0.03 & 1.06 $\pm$ 0.04 & $1.91^{+0.054}_{-0.093}$ & 15.3 & 0.627 $\pm$ 0.14 & $-$5.48 \\
094246.0+553102.7 & M8 & 16.88 $\pm$ 0.02 & 1.80 $\pm$ 0.02 & 3.96 $\pm$ 0.03 & 1.12 $\pm$ 0.03 & $2.81^{+0.12}_{-0.039}$ & 14.9 & 0.742 $\pm$ 0.14 & $-$5.58 \\
094720.0$-$002009.5 & M7 & 15.24 $\pm$ 0.01 & 1.20 $\pm$ 0.02 & 2.99 $\pm$ 0.02 & 0.91 $\pm$ 0.03 & $5.57^{+0.22}_{-0.66}$ & 14.2 & 5.73 $\pm$ 1.0 & $-$4.99 \\
094738.4+371016.5 & M7 & 15.10 $\pm$ 0.02 & 1.16 $\pm$ 0.02 & 2.91 $\pm$ 0.03 & 0.85 $\pm$ 0.03 & $5.28^{+0.31}_{-0.30}$ & 14.2 & 6.72 $\pm$ 1.3 & $-$4.90 \\
095246.2+062041.0 & M8 & 16.02 $\pm$ 0.02 & 1.54 $\pm$ 0.02 & 3.57 $\pm$ 0.03 & 0.99 $\pm$ 0.03 & $4.04^{+0.19}_{-0.74}$ & 14.5 & 2.16 $\pm$ 0.57 & $-$5.27 \\
101100.2+424502.9 & M8 & 17.06 $\pm$ 0.02 & 1.62 $\pm$ 0.03 & 3.70 $\pm$ 0.03 & 1.05 $\pm$ 0.03 & $1.85^{+0.048}_{-0.053}$ & 15.4 & 0.677 $\pm$ 0.18 & $-$5.44 \\
101707.5+130839.3 & L2 & 18.54 $\pm$ 0.02 & 1.79 $\pm$ 0.03 & 4.44 $\pm$ 0.03 & 1.39 $\pm$ 0.03 & $0.956^{+0.18}_{-0.016}$ & 16.1 & 0.201 $\pm$ 0.037 & $-$5.68 \\
104842.8+011158.0 & L1 & 17.25 $\pm$ 0.02 & 1.81 $\pm$ 0.03 & 4.33 $\pm$ 0.03 & 1.30 $\pm$ 0.03 & $2.91^{+0.18}_{-0.022}$ & 14.9 & 0.655 $\pm$ 0.11 & $-$5.65 \\
105029.7+615954.2 & M7 & 15.67 $\pm$ 0.01 & 1.13 $\pm$ 0.02 & 2.84 $\pm$ 0.02 & 0.93 $\pm$ 0.03 & $3.07^{+0.31}_{-0.13}$ & 14.8 & 4.38 $\pm$ 0.71 & $-$4.85 \\
105032.1+490741.4 & M8 & 16.47 $\pm$ 0.02 & 1.44 $\pm$ 0.03 & 3.49 $\pm$ 0.03 & 1.02 $\pm$ 0.03 & $2.64^{+0.070}_{-0.11}$ & 15.0 & 1.36 $\pm$ 0.28 & $-$5.29 \\
105547.2+080842.6 & M9 & 16.60 $\pm$ 0.02 & 1.69 $\pm$ 0.03 & 4.05 $\pm$ 0.03 & 1.18 $\pm$ 0.03 & $3.97^{+0.064}_{-0.18}$ & 14.5 & 0.991 $\pm$ 0.21 & $-$5.60 \\
110009.6+495746.5 & L3 & 20.09 $\pm$ 0.03 & 2.09 $\pm$ 0.03 & 4.81 $\pm$ 0.05 & 1.81 $\pm$ 0.05 & $0.434^{+0.0018}_{-0.096}$ & 16.9 & \nodata & \nodata \\
110153.8+341017.1 & M7 & 15.46 $\pm$ 0.01 & 1.06 $\pm$ 0.02 & 2.69 $\pm$ 0.02 & 0.79 $\pm$ 0.03 & $3.16^{+0.13}_{-0.15}$ & 14.8 & 5.32 $\pm$ 0.86 & $-$4.77 \\
110401.2+195922.3 & L5 & 19.35 $\pm$ 0.02 & 2.15 $\pm$ 0.03 & 4.97 $\pm$ 0.03 & 1.43 $\pm$ 0.04 & $0.760^{+0.18}_{-0.16}$ & 16.3 & 0.117 $\pm$ 0.041 & $-$5.81 \\
112329.3+015404.0 & M7 & 15.72 $\pm$ 0.02 & 1.53 $\pm$ 0.03 & 3.35 $\pm$ 0.04 & 1.01 $\pm$ 0.04 & $4.60^{+0.26}_{-1.0}$ & 14.4 & 3.45 $\pm$ 0.77 & $-$5.13 \\
113911.0+084112.0 & M8 & 16.84 $\pm$ 0.02 & 1.71 $\pm$ 0.03 & 3.92 $\pm$ 0.03 & 1.09 $\pm$ 0.03 & $2.77^{+0.093}_{-0.040}$ & 14.9 & 0.785 $\pm$ 0.18 & $-$5.55 \\
115242.4+243808.0 & M9 & 17.13 $\pm$ 0.01 & 1.80 $\pm$ 0.02 & 4.10 $\pm$ 0.02 & 1.25 $\pm$ 0.03 & $2.76^{+0.078}_{-0.10}$ & 14.9 & 0.651 $\pm$ 0.11 & $-$5.63 \\
115250.0+643252.4 & L2 & 20.01 $\pm$ 0.03 & 1.86 $\pm$ 0.04 & 4.41 $\pm$ 0.06 & 1.51 $\pm$ 0.08 & $0.268^{+0.012}_{-0.022}$ & 17.4 & \nodata & \nodata \\
120032.8+204852.7 & M8 & 16.54 $\pm$ 0.02 & 1.55 $\pm$ 0.03 & 3.68 $\pm$ 0.02 & 1.00 $\pm$ 0.03 & $2.87^{+0.12}_{-0.046}$ & 14.9 & 1.09 $\pm$ 0.28 & $-$5.42 \\
120133.5+404645.6 & M7 & 15.80 $\pm$ 0.01 & 1.27 $\pm$ 0.02 & 3.16 $\pm$ 0.02 & 0.90 $\pm$ 0.02 & $3.58^{+0.072}_{-0.47}$ & 14.6 & 3.00 $\pm$ 0.63 & $-$5.08 \\
121754.5+580135.0 & M7 & 16.86 $\pm$ 0.01 & 1.36 $\pm$ 0.02 & 3.23 $\pm$ 0.03 & 0.94 $\pm$ 0.04 & $1.44^{+0.15}_{-0.092}$ & 15.6 & 1.04 $\pm$ 0.23 & $-$5.14 \\
122127.7+025719.8 & L0 & 17.41 $\pm$ 0.02 & 1.80 $\pm$ 0.02 & 4.25 $\pm$ 0.03 & 1.22 $\pm$ 0.03 & $2.27^{+0.16}_{-0.034}$ & 15.1 & 0.518 $\pm$ 0.072 & $-$5.64 \\
123121.3+495923.3 & L3 & 19.23 $\pm$ 0.02 & 1.95 $\pm$ 0.03 & 4.61 $\pm$ 0.04 & 1.48 $\pm$ 0.04 & $0.696^{+0.044}_{-0.066}$ & 16.4 & 0.116 $\pm$ 0.043 & $-$5.78 \\
123323.3+311631.2 & M7 & 16.07 $\pm$ 0.01 & 1.24 $\pm$ 0.02 & 3.01 $\pm$ 0.02 & 0.97 $\pm$ 0.03 & $2.55^{+0.25}_{-0.10}$ & 15.0 & 2.65 $\pm$ 0.50 & $-$4.98 \\
123927.3+551537.3 & L5 & 19.64 $\pm$ 0.04 & 2.16 $\pm$ 0.05 & 4.93 $\pm$ 0.05 & 1.92 $\pm$ 0.04 & $0.891^{+0.011}_{-0.12}$ & 16.1 & 0.0896 $\pm$ 0.057 & $-$6.00 \\
125041.1+402134.7 & M7 & 15.85 $\pm$ 0.01 & 1.28 $\pm$ 0.01 & 3.10 $\pm$ 0.02 & 0.97 $\pm$ 0.02 & $3.10^{+0.34}_{-0.11}$ & 14.8 & 3.10 $\pm$ 0.57 & $-$5.00 \\
125222.6+025205.6 & M8 & 16.13 $\pm$ 0.02 & 1.57 $\pm$ 0.02 & 3.63 $\pm$ 0.03 & 0.99 $\pm$ 0.03 & $4.23^{+0.14}_{-0.13}$ & 14.5 & 1.80 $\pm$ 0.38 & $-$5.37 \\
125312.4+403400.5 & M7 & 15.58 $\pm$ 0.01 & 1.41 $\pm$ 0.02 & 3.40 $\pm$ 0.02 & 1.02 $\pm$ 0.02 & $5.47^{+0.28}_{-0.20}$ & 14.2 & 3.29 $\pm$ 0.72 & $-$5.22 \\
125737.2$-$011336.2 & L5 & 20.76 $\pm$ 0.05 & 2.21 $\pm$ 0.06 & 4.82 $\pm$ 0.10 & 1.82 $\pm$ 0.11 & $0.250^{+0.0059}_{-0.065}$ & 17.5 & \nodata & \nodata \\
130407.4+403615.7 & M9 & 16.89 $\pm$ 0.02 & 1.47 $\pm$ 0.03 & 3.71 $\pm$ 0.03 & 1.09 $\pm$ 0.03 & $2.23^{+0.042}_{-0.11}$ & 15.1 & 0.851 $\pm$ 0.18 & $-$5.42 \\
131711.5+184923.0 & M9 & 16.27 $\pm$ 0.01 & 1.62 $\pm$ 0.02 & 3.84 $\pm$ 0.02 & 1.10 $\pm$ 0.03 & $4.34^{+0.20}_{-0.18}$ & 14.4 & 1.66 $\pm$ 0.34 & $-$5.42 \\
133650.5+475132.1 & M8 & 16.15 $\pm$ 0.02 & 1.53 $\pm$ 0.02 & 3.49 $\pm$ 0.03 & 0.99 $\pm$ 0.03 & $3.62^{+0.15}_{-0.11}$ & 14.6 & 1.71 $\pm$ 0.41 & $-$5.33 \\
134316.5+394509.2 & L5 & 21.30 $\pm$ 0.08 & 2.36 $\pm$ 0.09 & 5.14 $\pm$ 0.11 & 2.01 $\pm$ 0.08 & $0.265^{+0.0016}_{-0.023}$ & 17.5 & \nodata & \nodata \\
134607.4+084234.5 & L2 & 20.02 $\pm$ 0.03 & 1.86 $\pm$ 0.04 & 4.27 $\pm$ 0.08 & 1.58 $\pm$ 0.10 & $0.254^{+0.016}_{-0.018}$ & 17.5 & \nodata & \nodata \\
140322.0+300754.8 & M9 & 16.54 $\pm$ 0.02 & 1.72 $\pm$ 0.03 & 3.86 $\pm$ 0.03 & 1.08 $\pm$ 0.03 & $3.45^{+0.16}_{-0.041}$ & 14.7 & \nodata & \nodata \\
141351.7$-$002208.5 & M7 & 16.86 $\pm$ 0.02 & 1.22 $\pm$ 0.03 & 2.88 $\pm$ 0.03 & 0.84 $\pm$ 0.05 & $1.15^{+0.075}_{-0.11}$ & 15.9 & 1.23 $\pm$ 0.26 & $-$4.97 \\
142224.2+211607.5 & M8 & 16.08 $\pm$ 0.01 & 1.64 $\pm$ 0.02 & 3.65 $\pm$ 0.03 & 0.99 $\pm$ 0.03 & $4.37^{+0.12}_{-0.48}$ & 14.4 & 1.83 $\pm$ 0.41 & $-$5.38 \\
142257.1+082752.1 & L2 & 19.47 $\pm$ 0.02 & 1.79 $\pm$ 0.03 & 4.37 $\pm$ 0.05 & 1.45 $\pm$ 0.07 & $0.419^{+0.022}_{-0.025}$ & 17.0 & 0.108 $\pm$ 0.054 & $-$5.59 \\
142631.6+155700.9 & M9 & 16.95 $\pm$ 0.02 & 1.71 $\pm$ 0.03 & 4.04 $\pm$ 0.03 & 1.18 $\pm$ 0.02 & $2.87^{+0.15}_{-0.041}$ & 14.9 & 0.724 $\pm$ 0.15 & $-$5.60 \\
142831.2+592335.3 & L5 & 19.66 $\pm$ 0.03 & 2.01 $\pm$ 0.04 & 4.88 $\pm$ 0.05 & 1.52 $\pm$ 0.04 & $0.561^{+0.062}_{-0.077}$ & 16.6 & 0.0816 $\pm$ 0.051 & $-$5.84 \\
143043.5+291541.3 & L2 & 18.76 $\pm$ 0.02 & 1.81 $\pm$ 0.03 & 4.48 $\pm$ 0.03 & 1.50 $\pm$ 0.03 & $0.892^{+0.081}_{-0.018}$ & 16.1 & 0.166 $\pm$ 0.048 & $-$5.73 \\
143130.7+143653.4 & L3 & 19.71 $\pm$ 0.03 & 2.11 $\pm$ 0.03 & 4.56 $\pm$ 0.05 & 1.03 $\pm$ 0.07 & $0.339^{+0.020}_{-0.0079}$ & 17.2 & \nodata & \nodata \\
143609.7+290035.5 & M8 & 17.13 $\pm$ 0.02 & 1.65 $\pm$ 0.03 & 3.84 $\pm$ 0.03 & 1.13 $\pm$ 0.02 & $1.99^{+0.075}_{-0.024}$ & 15.3 & 0.602 $\pm$ 0.14 & $-$5.52 \\
144013.0+631908.1 & M7 & 15.98 $\pm$ 0.02 & 1.14 $\pm$ 0.02 & 2.86 $\pm$ 0.02 & 0.95 $\pm$ 0.02 & $2.41^{+0.23}_{-0.10}$ & 15.1 & 2.68 $\pm$ 0.42 & $-$4.95 \\
144022.8+133920.8 & M7 & 15.77 $\pm$ 0.01 & 1.44 $\pm$ 0.02 & 3.38 $\pm$ 0.02 & 1.06 $\pm$ 0.03 & $4.95^{+0.17}_{-0.82}$ & 14.3 & 2.99 $\pm$ 0.56 & $-$5.22 \\
144825.6+103158.8 & L5 & 19.63 $\pm$ 0.02 & 2.17 $\pm$ 0.03 & 5.07 $\pm$ 0.04 & 1.87 $\pm$ 0.04 & $0.921^{+0.056}_{-0.064}$ & 16.1 & 0.0938 $\pm$ 0.069 & $-$5.99 \\
144834.1+104805.5 & M7 & 16.01 $\pm$ 0.02 & 1.24 $\pm$ 0.03 & 3.09 $\pm$ 0.03 & 0.89 $\pm$ 0.04 & $2.77^{+0.10}_{-0.27}$ & 14.9 & 2.76 $\pm$ 0.53 & $-$5.00 \\
145201.3+093136.8 & L3 & 19.67 $\pm$ 0.03 & 1.89 $\pm$ 0.03 & 4.25 $\pm$ 0.07 & 1.17 $\pm$ 0.11 & $0.281^{+0.015}_{-0.0045}$ & 17.4 & \nodata & \nodata \\
145237.2+632319.7 & M8 & 16.31 $\pm$ 0.02 & 1.35 $\pm$ 0.03 & 3.31 $\pm$ 0.04 & 1.04 $\pm$ 0.04 & $2.75^{+0.072}_{-0.27}$ & 14.9 & 1.82 $\pm$ 0.38 & $-$5.18 \\
145348.4+373316.9 & M8 & 16.83 $\pm$ 0.02 & 1.62 $\pm$ 0.03 & 3.66 $\pm$ 0.03 & 1.05 $\pm$ 0.03 & $2.26^{+0.054}_{-0.15}$ & 15.1 & 0.877 $\pm$ 0.20 & $-$5.41 \\
145824.4+283955.8 & M9 & 17.11 $\pm$ 0.02 & 1.78 $\pm$ 0.02 & 4.03 $\pm$ 0.02 & 1.23 $\pm$ 0.03 & $2.46^{+0.12}_{-0.041}$ & 15.0 & 0.630 $\pm$ 0.13 & $-$5.59 \\
145919.3+060453.6 & M8 & 16.60 $\pm$ 0.02 & 1.48 $\pm$ 0.03 & 3.44 $\pm$ 0.03 & 0.96 $\pm$ 0.03 & $2.26^{+0.050}_{-0.23}$ & 15.1 & 1.24 $\pm$ 0.32 & $-$5.26 \\
150026.3$-$003927.9 & M7 & 14.56 $\pm$ 0.01 & 1.22 $\pm$ 0.02 & 2.96 $\pm$ 0.03 & 0.95 $\pm$ 0.03 & $9.26^{+0.86}_{-0.28}$ & 13.6 & 11.3 $\pm$ 1.8 & $-$4.92 \\
150048.8$-$013142.6 & M7 & 15.82 $\pm$ 0.02 & 1.27 $\pm$ 0.03 & 3.02 $\pm$ 0.03 & 0.93 $\pm$ 0.03 & $3.28^{+0.11}_{-0.33}$ & 14.7 & 3.15 $\pm$ 0.56 & $-$5.02 \\
150108.1+225001.7 & M9 & 16.08 $\pm$ 0.02 & 1.79 $\pm$ 0.02 & 4.22 $\pm$ 0.02 & 1.16 $\pm$ 0.03 & $7.48^{+0.71}_{-0.14}$ & 13.8 & 1.70 $\pm$ 0.21 & $-$5.64 \\
151727.7+335702.4 & M7 & 15.63 $\pm$ 0.01 & 1.10 $\pm$ 0.02 & 2.87 $\pm$ 0.02 & 0.98 $\pm$ 0.03 & $3.33^{+0.32}_{-0.12}$ & 14.7 & 4.19 $\pm$ 0.72 & $-$4.90 \\
153154.6+424331.3 & M7 & 16.96 $\pm$ 0.01 & 1.40 $\pm$ 0.03 & 3.18 $\pm$ 0.03 & 0.99 $\pm$ 0.04 & $1.39^{+0.082}_{-0.081}$ & 15.7 & 1.06 $\pm$ 0.23 & $-$5.12 \\
153453.3+121949.4 & L6 & 20.43 $\pm$ 0.04 & 2.29 $\pm$ 0.05 & 5.10 $\pm$ 0.06 & 1.51 $\pm$ 0.06 & $0.393^{+0.0059}_{-0.077}$ & 17.0 & 0.0482 $\pm$ 0.026 & $-$5.91 \\
153659.8+585346.0 & M7 & 16.35 $\pm$ 0.01 & 1.12 $\pm$ 0.02 & 2.81 $\pm$ 0.03 & 0.95 $\pm$ 0.03 & $1.59^{+0.093}_{-0.067}$ & 15.5 & 2.37 $\pm$ 0.36 & $-$4.83 \\
154420.4$-$002326.3 & M7 & 15.78 $\pm$ 0.01 & 1.24 $\pm$ 0.02 & 3.08 $\pm$ 0.02 & 0.99 $\pm$ 0.03 & $3.22^{+0.44}_{-0.13}$ & 14.7 & 3.96 $\pm$ 0.60 & $-$4.91 \\
154455.1+330145.2 & L6 & 20.52 $\pm$ 0.05 & 2.21 $\pm$ 0.06 & 4.97 $\pm$ 0.07 & 1.61 $\pm$ 0.07 & $0.321^{+0.0052}_{-0.071}$ & 17.3 & \nodata & \nodata \\
154849.0+172235.3 & L8 & 21.34 $\pm$ 0.08 & 2.50 $\pm$ 0.08 & 5.23 $\pm$ 0.13 & 1.65 $\pm$ 0.13 & $0.193^{+0.017}_{-0.051}$ & 17.8 & \nodata & \nodata \\
155008.5+145517.0 & L3 & 19.58 $\pm$ 0.02 & 1.88 $\pm$ 0.03 & 4.81 $\pm$ 0.05 & 1.52 $\pm$ 0.05 & $0.596^{+0.0041}_{-0.14}$ & 16.6 & 0.0829 $\pm$ 0.030 & $-$5.86 \\
155038.1+304104.2 & M7 & 16.42 $\pm$ 0.01 & 1.46 $\pm$ 0.02 & 3.42 $\pm$ 0.02 & 1.05 $\pm$ 0.02 & $2.63^{+0.16}_{-0.11}$ & 15.0 & 1.64 $\pm$ 0.31 & $-$5.20 \\
155120.8+432930.3 & L3 & 19.58 $\pm$ 0.02 & 1.87 $\pm$ 0.03 & 4.45 $\pm$ 0.05 & 1.51 $\pm$ 0.06 & $0.458^{+0.045}_{-0.060}$ & 16.9 & \nodata & \nodata \\
155259.0+294848.3 & L0 & 18.11 $\pm$ 0.02 & 1.89 $\pm$ 0.02 & 4.63 $\pm$ 0.03 & 1.46 $\pm$ 0.03 & $1.68^{+0.46}_{-0.056}$ & 15.5 & 0.293 $\pm$ 0.053 & $-$5.76 \\
161345.5+170827.4 & M9 & 17.55 $\pm$ 0.01 & 1.75 $\pm$ 0.02 & 4.08 $\pm$ 0.02 & 1.28 $\pm$ 0.03 & $1.81^{+0.15}_{-0.024}$ & 15.4 & 0.437 $\pm$ 0.071 & $-$5.62 \\
161544.1+355858.4 & L4 & 19.20 $\pm$ 0.02 & 1.90 $\pm$ 0.03 & 4.66 $\pm$ 0.04 & 1.60 $\pm$ 0.04 & $0.776^{+0.13}_{-0.12}$ & 16.3 & 0.0901 $\pm$ 0.053 & $-$5.93 \\
161648.0+444303.4 & M7 & 15.36 $\pm$ 0.02 & 1.06 $\pm$ 0.03 & 2.70 $\pm$ 0.03 & 0.90 $\pm$ 0.03 & $3.32^{+0.18}_{-0.22}$ & 14.7 & 6.11 $\pm$ 0.90 & $-$4.73 \\
161731.0+514022.4 & M8 & 17.54 $\pm$ 0.01 & 1.52 $\pm$ 0.02 & 3.64 $\pm$ 0.02 & 1.11 $\pm$ 0.03 & $1.15^{+0.028}_{-0.039}$ & 15.9 & 0.485 $\pm$ 0.12 & $-$5.38 \\
162338.8+161554.5 & M7 & 15.77 $\pm$ 0.02 & 1.15 $\pm$ 0.02 & 3.01 $\pm$ 0.03 & 0.99 $\pm$ 0.03 & $3.14^{+0.27}_{-0.11}$ & 14.8 & 3.96 $\pm$ 0.56 & $-$4.90 \\
162718.2+353835.6 & M7 & 15.44 $\pm$ 0.01 & 1.35 $\pm$ 0.03 & 3.09 $\pm$ 0.02 & 0.99 $\pm$ 0.03 & $5.25^{+0.12}_{-0.74}$ & 14.2 & 4.71 $\pm$ 0.83 & $-$5.05 \\
163112.2+322711.4 & M8 & 16.54 $\pm$ 0.01 & 1.56 $\pm$ 0.03 & 3.38 $\pm$ 0.02 & 0.91 $\pm$ 0.03 & $2.27^{+0.11}_{-0.13}$ & 15.1 & 1.21 $\pm$ 0.32 & $-$5.27 \\
164438.1+284614.5 & L2 & 19.91 $\pm$ 0.03 & 1.78 $\pm$ 0.04 & 4.26 $\pm$ 0.07 & 1.48 $\pm$ 0.10 & $0.275^{+0.00036}_{-0.025}$ & 17.4 & \nodata & \nodata
\enddata
\tablenotetext{a}{Apparent bolometric magnitude; adopting L$\odot = 3.8265 \times 10^{33} $ergs s$^{-1}$ \citep{Kopp2011} and $M_{bol,\odot}$=4.7554 then $m_{bol}$ = $-2.5 \times \log_{10}(F_{bol}) -  11.482$}
\end{deluxetable*}

\clearpage

\end{document}